\documentclass[12pt,preprint]{aastex}

\shorttitle{The ACS Virgo Cluster Survey. I. Introduction}
\shortauthors{C\^ot\'e \etal}

\begin{document}


\title{The ACS Virgo Cluster Survey. I. Introduction to the Survey\altaffilmark{1}}


\author{Patrick C\^ot\'e\altaffilmark{2},
John P. Blakeslee\altaffilmark{3},
Laura Ferrarese\altaffilmark{2},
Andr\'es Jord\'an\altaffilmark{2,4},
Simona Mei\altaffilmark{5},
David Merritt\altaffilmark{6},
Milo\v s Milosavljevi\'c\altaffilmark{7,8},
Eric W. Peng\altaffilmark{2},
John L. Tonry\altaffilmark{9},
Michael J. West\altaffilmark{10}}


\altaffiltext{1}{Based on observations with the NASA/ESA {\it Hubble Space Telescope}
obtained at the Space Telescope Science Institute, which is operated by the association 
of Universities for Research in Astronomy, Inc., under NASA contract NAS 5-26555.}
\altaffiltext{2}{Department of Physics and Astronomy, Rutgers University, New Brunswick, 
NJ 08854; pcote@physics.rutgers.edu, lff@physics.rutgers.edu, andresj@physics.rutgers.edu,
ericpeng@physics.rutgers.edu}
\altaffiltext{3}{Department of Physics and Astronomy, The Johns Hopkins University, 
3400 North Charles Street, Baltimore, MD 21218-2686; jpb@pha.jhu.edu}
\altaffiltext{4}{Claudio Anguita Fellow}
\altaffiltext{5}{Institut d'Astrophysique Spatiale, Universit\'e Paris-Sud, B\^at. 121, 
91405 Orsay, France; simona.mei@ias.u-psud.fr}
\altaffiltext{6}{Department of Physics, Rochester Institute of Technology,
84 Lomb Memorial Drive, Rochester, NY 14623; merritt@mail.rit.edu}
\altaffiltext{7}{Theoretical Astrophysics, California Institute of Technology, Mail Stop 
130-33, Pasadena, CA 91125; milos@tapir.caltech.edu}
\altaffiltext{8}{Sherman M. Fairchild Fellow}
\altaffiltext{9}{Institute for Astronomy, University of Hawaii, 2680 Woodlawn Drive,
Honolulu, HI 96822; jt@ifa.hawaii.edu}
\altaffiltext{10}{Department of Physics and Astronomy, University of Hawaii,
Hilo, HI 96720; westm@hawaii.edu}


\slugcomment{Accepted to the {\it Astrophysical Journal Supplement Series}}

\begin{abstract}
The Virgo Cluster is the dominant mass concentration in the Local Supercluster and
the largest collection of elliptical and lenticular galaxies in the nearby universe.
In this paper, we present an introduction to the ACS Virgo Cluster Survey: a 
program to image, in the F475W and F850LP bandpasses
($\approx$ Sloan $g$ and $z$),
100 early-type galaxies in the Virgo Cluster using the {\it Advanced Camera for
Surveys} on the {\it Hubble Space Telescope}. We describe the selection of
the program galaxies and their ensemble properties, the choice of filters, the
field placement and orientation, the limiting magnitudes of the survey, 
coordinated parallel observations of 100 ``intergalactic" fields
with WFPC2, and supporting ground-based spectroscopic observations of
the program galaxies. In terms of depth, spatial resolution, sample
size and homogeneity, this represents the most comprehensive imaging survey to
date of early-type galaxies in a cluster environment. We briefly describe the
main scientific goals of the survey which include the measurement of 
luminosities, metallicities, ages, and structural parameters for the
many thousands of globular clusters associated with these galaxies, a
high-resolution isophotal analysis of galaxies spanning a factor of
$\sim$ 450 in luminosity and sharing a common environment,
the measurement of accurate distances for the full sample of galaxies using the
method of surface brightness fluctuations, and a determination of the 
three-dimensional structure of Virgo itself.
\end{abstract}




\section{Introduction}
\label{sec:introduction}

Clusters of galaxies are among the largest gravitationally bound structures
in the universe. Although they contain a small fraction of all galaxies
in the universe ($e.g.$, Turner \& Tyson 1999), they
have long provided invaluable constraints on structure formation and galaxy
evolution. During the past few decades, it has become abundantly clear that
the formation of galaxy clusters is a highly protracted process: most 
clusters show some degree of substructure (Geller \& Beers 1982;
Dressler \& Shectman 1988;  Fitchett \& Merritt 1988; Mohr, Fabricant \&
Geller 1993; Bird 1994)
and there is mounting evidence that, even among nearby clusters, the
process of virialization may not be fully complete ($e.g.$, Adami,
Biviano \& Mazure 1998; Drinkwater, Gregg \& Colless 2001).

Rich clusters may contain several thousands of member galaxies within a 
region $\sim$ 1 Mpc in diameter. Collectively, these galaxies usually
comprise only a few percent of the total gravitating mass 
($e.g.$, Carlberg et al. 1996). Instead, cluster masses are dominated
by dark matter, and to a lesser extent, by the X-ray emitting gas which
fills their gravitational potential wells (typically $\gtrsim 90$\% and
10--30\% of the total mass, respectively; B\"ohringer 1995).
Yet despite their relatively minor contribution to the cluster mass budget,
and to that of the universe in general,
much of our understanding of galaxy formation and evolution
is based on observations of galaxies in cluster environments.

Virgo is probably the most thoroughly studied cluster of
galaxies, being the cluster nearest to our own Galaxy. In most
respects, its properties are typical of large clusters. 
It is a relatively populous system (Abell richness class I;
Girardi et al. 1995), consisting of $\approx$ 2000 cataloged members
brighter than $B_T = 18$
(Binggeli, Sandage \& Tammann 1987). Mass estimates vary substantially
($e.g.$, 0.15--1.5$\times10^{15}M_{\odot}$; B\"ohringer et al. 1994;
Schindler, Binggeli \& B\"ohringer 1999; McLaughlin 1999a; Tonry et al. 2000;
Fouqu\'e et al.  2001), but generally fall within the range typical
of rich clusters
(B\"ohringer 1995). It contains vast quantities of X-ray emitting gas
(Takano et al. 1989; B\"ohringer et al. 1994;  Schindler et al.
1999), and shows clear evidence for both substructure and 
non-virialized motions ($e.g.,$ Huchra 1985; Binggeli et~al. 1987;
Binggeli, Popescu \& Tammann 1993). Indeed, 
the property which sets Virgo aside from other clusters
is proximity --- at a distance of $\approx 17$ Mpc 
(Tonry et al. 2001), it may be studied in a level of detail which
will never be possible with more distant systems.

The discovery of the cluster itself dates back to Messier and M\'echain,
who noted that 13 of the entries in Messier's catalog lie in the constellation of
Virgo (Messier 1781).\footnote{Messier writes ``The
constellation Virgo and especially the northern wing is one of the
constellations which encloses the most nebulae. This catalog contains 13
which have been determined, viz. Nos. 49, 58, 59, 60, 61, 84, 85, 86, 87,
88, 89, 90 and 91.... Most of these nebulae have been pointed out to me by
M. M\'echain." It is now recognized that three additional galaxies from
Messier's catalog (M98, M99 and M100) are also Virgo members.}
The first systematic studies of the cluster were those
of Shapley \& Ames (1926; 1929a-e), Hubble \& Humason (1931) and Smith (1936).
Subsequent studies by Zwicky (1942; 1957) and Holmberg (1958) produced both a
better understanding of its spatial structure and a more complete census of
cluster members. The cluster richness and its large dynamical mass led
de Vaucouleurs (1956; 1961) to identify it as the nucleus of
the Local Supercluster. Surveys of the Virgo Cluster culminated
in the 1980s, with the heroic efforts of Bruno Binggeli, Allan Sandage
and Gustav Tammann ($e.g.,$ Binggeli, Sandage \& Tammann 1984; 1985; 1987;
Sandage, Binggeli \& Tammann 1985). Their Virgo Cluster Catalog (VCC)
of 2096 galaxies, based on deep photographic images covering an area of
$\approx$ 140 deg$^{2}$, still stands as the most complete census of Virgo
galaxies brighter than $B_T = 18$.

Continuing studies of Virgo and its constituent galaxies at
X-ray (Schindler et al. 1999; B\"ohringer et al. 2001; Allen et al. 2001),
ultraviolet (Lieu et al. 1996; Cortese et al. 2003),
optical (Trentham \& Hodgkin 2002; Sabatini et al. 2003),
infrared (Boselli et al. 1997; Tuffs et al. 2002), and
radio (van Driel et al. 2000; Solanes et al. 2002) wavelengths ---
as well as combinations thereof (Gavazzi et al. 2002; 2003; Boselli,
Gavazzi \& Sanvito 2003) --- have largely confirmed the view of Virgo as
an environment that is, in most respects, typical of rich clusters
(see, $e.g.$, Binggeli 1999 and references therein).
It therefore seems likely that Virgo will continue to play a central role
in the study of galaxy formation and evolution. This is particularly
true in the case of early-type galaxies since Virgo contains, by
far, the largest collection of early-type giant and dwarf galaxies
in the local universe.

Not surprisingly, this enormous reservoir of galaxies has been a favorite
target of the {\it Hubble Space Telescope} (HST) since its launch in
1990. To date, thousands of high-resolution images for fields in Virgo have
been obtained with FOC, WFPC, NICMOS, STIS and, especially, WFPC2. The
scientific output from the first decade of Virgo research with HST
touches upon a remarkably diverse range of
astrophysical topics including the
extragalactic distance scale, 
accreting and pulsating stars,
the cores of spiral and early-type galaxies,
the stellar populations of galaxies, 
extragalactic globular cluster systems,
dwarf galaxies,
the nuclear and circumstellar structure of active galaxies,
intergalactic stars and diffuse light,
the ultraviolet/optical/IR morphologies of nearby galaxies,
galaxy collisions and interactions, and
pre- and post-main-sequence stages of stellar evolution.

With the installation of the {\it Advanced Camera for Surveys} (ACS; 
Ford et al. 1998) during Servicing Mission 3B in March 2002, the imaging
capabilities of HST have improved dramatically ($i.e.$, by
nearly an order of magnitude in discovery efficiency).
Relative to WFPC2, the Wide Field Channel (WFC)
mode on ACS boasts a factor of $\sim$ two increase in 
imaging area and a typical gain
of 3--4 times in throughput over the wavelength range 0.4--1.0~$\mu$m.
ACS thus opens the possibility of obtaining deep, high-resolution
optical images for a large sample of Virgo Cluster galaxies in a
relatively modest allocation of observing time.

In this paper, the first in a series, we present an introduction to 
such a program: the ACS Virgo Cluster Survey. The program, which
was awarded 100 orbits in Cycle 11, aims to obtain deep
F475W ($\approx$ Sloan $g$) and F850LP ($\approx$ Sloan $z$)
ACS/WFC images for 100 early-type galaxies that are confirmed members
of the Virgo Cluster. The paper is organized as follows: we present a
brief description of the primary scientific goals of the survey and
the anticipated data products in \S\ref{sec:motivation},
the selection of sample galaxies in \S\ref{sec:properties}, and
the adopted observing strategy, including choice of filters,
exposure times, limiting magnitudes and field placement, in
\S\ref{sec:strategy}. Since the survey also includes supplemental WFPC2
parallel observations of 100 intergalactic fields in Virgo and ground-based, long-slit
spectroscopy of the program galaxies, brief descriptions of these
observations are given in \S\ref{sec:parallel} and \S\ref{sec:spectroscopy},
respectively.
Scientific results from the survey will be presented in future papers
in this series.

\section{Motivation}

It is clear from the above discussion that high-spatial resolution,
multi-color imaging for a large sample of Virgo Cluster galaxies would
provide a valuable resource for many different scientific applications.
In this section, we give a brief description of three specific 
scientific programs which have motivated the present survey.

\label{sec:motivation}
\subsection{Extragalactic Globular Cluster Systems}

Globular clusters (GCs) are attractive diagnostics of galaxy formation and
evolution for several reasons. They are ubiquitous, found in association with
galaxies of all type spanning a factor of $\sim$~10$^4$ in mass. Roughly
0.25\% of the total baryonic mass in galaxies
($i.e.$, gas and stars) takes the form of GCs (McLaughlin 1999b), meaning
that the most luminous galaxies may contain many thousands of GCs
($e.g.$, Harris, Pritchet \& McClure 1995; Blakeslee et al. 2003). 
They are the brightest, readily identifiable objects in early-type galaxies,
and since they consist of coeval stars with nearly identical metallicities,
the interpretation of their broadband colors is much simpler than in
composite stellar systems such as their host galaxies. 

The metallicity distribution of a given GC system reflects
the chemical enrichment history of its parent galaxy (see, $e.g.$, the
review of West et al. 2004), while the measurement of GC ages provides
a means of reconstructing the likely complex history of
mergers and star formation undergone by the galaxy (Cohen, Blakeslee \&
Ryzhov 1998; Puzia et al. 1999; Jord\'an et al. 2002; Cohen,
Blakeslee \& C\^ot\'e 2003). 
Because GCs represent dynamical test particles orbiting in the
gravitational potential wells of their parent galaxies, it is possible
to measure both the GC orbital properties and the distribution of
dark matter from their radial velocities (Cohen \& Ryzhov 1997; C\^ot\'e et al. 2001;
2003). The radii and velocities of the GCs provide joint constraints on the form of
the gravitational potential and the orbital properties of the GCs
themselves (Keenan 1981; Innanen, Harris \& Webbink 1983; Bellazzini 2004),
while the peak of the GC luminosity function (GCLF) continues 
to receive attention as a possible standard candle (see the review of Harris 2001).
Finally, observations with the {\it Chandra X-Ray Observatory} have shown that many
early-type galaxies are surrounded by numerous low-mass X-ray binaries,
roughly half of which are associated with GCs ($e.g.$, Sarazin, Irwin \&
Bregman 2001; Angelini, Loewenstein \& Mushotzky 2001; Maccarone,
Kundu \& Zepf 2003; Jord\'an et al. 2004a, hereafter Paper III).

Given their versatility in addressing such a wide variety of astrophysical
problems, it is not surprising that considerable effort continues to be
devoted to the study of extragalactic GC systems. HST has played a pivotal role
in GC research during the past decade, with important contributions coming
from the analysis of archival WFPC2 observations for a
few dozen galaxies ($e.g.$, Gebhardt \& Kissler-Patig 1999; Kundu \&
Whitmore 2001; Larsen et al. 2001). For the most part, the images
used in these studies were planned with other scientific goals in mind, 
so that the choice of filters, field placement, exposure times, dither
patterns, etc., were not optimized for the study of GCs. Perhaps 
inevitably, such archival observations are somewhat heterogeneous
in nature, whereas certain scientific questions
(for example, the usefulness of the GCLF as a standard candle) are
better addressed with extremely homogeneous datasets. 

The ACS Virgo Cluster Survey has been designed to produce the largest,
most homogeneous catalog of extragalactic GCs yet assembled. Deep ACS
images in F475W and F850LP (see \S\ref{sec:strategy}) 
will probe the brightest $\sim$ 90\% of the GCLF in our
100 program galaxies, yielding a sample of $\approx$ 13,000 GCs.
For individual
GCs, measured properties include coordinates, magnitudes, colors,
metallicities, and structural parameters ($i.e.$, half-light radius and
King concentration index). The high sensitivity and high spatial resolution 
of HST/ACS are essential since the clusters are both faint and compact. For
instance, the GCLF has a apparently universal, roughly Gaussian form
(Harris 2001), meaning that the GCLFs of Virgo Cluster galaxies are
expected to peak near $g_{\rm 475} \sim 24.2$ and
$z_{\rm 850} \sim 23.0$.\footnote{All $g_{\rm 475}$ and $z_{\rm 850}$
magnitudes are on the AB system (Fukugita et al. 1996).}
At the distance of Virgo, 1$^{\prime\prime}$ corresponds to a projected
distance of 82$\pm$6 pc (Tonry et al. 2001), so the typical
half-light radius of Galactic GCs ($r_h \simeq 3$pc; van den Bergh
1994) translates to slightly less than one ACS/WFC pixel. Thus, 
while the measurement of GC structural parameters at the distance of
Virgo is certainly feasible ($e.g.$, Larsen et al. 2001; Paper III),
it is nevertheless challenging, even with excellent spatial
resolution afforded by HST.

Future papers in this series will examine the properties of GCs in 
the program galaxies. First results are presented in Paper III,
which examines the connection between
low-mass X-ray binaries and GCs in VCC1316 (= M87) using
the ACS images and ACIS observations from {\it Chandra}.

\subsection{The Inner Structure of Early-Type Galaxies}

Surface photometry and isophotal studies have long provided clues to the
origin and structure of early-type galaxies (see Kormendy \& Djorgovski
1989 for a review). Yet until the launch of HST, the central
regions of these galaxies remained entirely unexplored, being
unresolved in ground-based studies with seeing of
$\sim 1^{\prime\prime}$. The first, high-resolution glimpse into
inner regions of early-type galaxies was that of Jaffe et al.
(1994), who used WFPC to obtain the $V$-band images of 
14 E and E/S0 galaxies in Virgo. van den Bosch et al. (1994)
carried out an isophotal analysis for these galaxies, finding
evidence for both significant quantities of dust and embedded
stellar disks. Major and minor-axis surface brightness profiles for
these galaxies were presented by Ferrarese et al. (1994),
who classified them into two categories based on the
slope of their central brightness profiles. Similar conclusions
were later reached by Lauer et al. (1995) using $V$-band HST
images for 45 early-type galaxies with distances in the range
$3 - 280$~Mpc.

The discovery of gas, dust and embedded stellar disks in the 
cores of these galaxies (see Tran et al. 2001 and references 
therein) has prompted a rethinking of the traditional
view of early-type galaxies as smooth, uniformly-old stellar
systems. It is now well established that both the central surface
brightness profile and the isophotal structure of a given galaxy
reflect its merger history and/or the presence of a
central supermassive black hole (Ebisuzaki et al.  1991;
Quinlan \& Hernquist 1997; Faber et al. 1997). The first attempts
to model quantitatively the evolution in central surface brightness 
during galaxy mergers that involve supermassive black hole binaries
have been presented in Milosavljevi\'c \& Merritt (2001).
Using structural parameters derived
from HST imaging, Ravindranath, Ho \& Filippenko (2002) and 
Milosavljevi\'c et al. (2002) showed 
the predictions of these binary black hole merger models are
broadly consistent with observations.

However, there are reasons to believe that the picture is
far from complete.
Recent suggestions that the manifold of early-type
galaxies can be neatly divided into two families according to the
central slope, $\gamma$, of their luminosity profiles ($i.e.$,
``core" versus ``power law" galaxies) have been called into question.
For instance, Rest et al. (2001) present results from the largest
imaging survey of early-type galaxies conducted to date, consisting
of $R$-band WFPC2 images for 67 galaxies over the range
$6 - 54$~Mpc. They
find a class of intermediate luminosity galaxies
whose luminosity profiles cannot be classified unambiguously as either
cores or power laws. Similar conclusions were found by
Ravindranath et~al. (2001), who used NICMOS imaging to 
examine the brightness profiles of Lauer et~al. (1995)
in the $H$-band, where the effects of dust extinction are minimized.
In addition, Graham et al. (2003) have examined in detail the
so-called ``Nuker Law" parameterization used by Lauer et al. (1995)
to model the brightness profiles of their galaxies. This particular
model provides an adequate representation of galaxy brightness
profiles having limited radial extent (i.e., the WFPC2 brightness
profiles published by Lauer et al. typically extent to
$\lesssim 10^{\prime\prime}$), but fails to reproduce the curvature
exhibited by the surface brightness profiles of real galaxies 
over $\sim$ two or more decades in  radius. 
Thus, Graham et al. (2003) find that the very classification
of certain galaxies as power laws may be inappropriate, and argue
that generalized Sersic models better represent
the brightness profiles of real galaxies.

Virgo contains an enormous collection of early-type galaxies which, lying at
a nearly identical distance and spanning a wide range in luminosity and mass,
offer attractive targets for imaging with ACS. At a distance of
only $\sim$ 17~Mpc, structures with physical sizes of $\sim$~10~pc
are resolved directly with ACS/WFC.
In addition, ACS/WFC enjoys a tremendous increase in field size relative
to WFPC2, so it is possible, at least for the brightest galaxies, to
measure brightness profiles
out to radii of $\sim$~140$^{\prime\prime}$ ($\approx$~11~kpc) in a single
pointing.  And since the galaxies themselves lie nearly at the same
distance (see \S\ref{sec:distance}), distance errors and biases due to
varying spatial sampling are greatly reduced.

In future papers in this series, we will present F475W and F850LP 
luminosity profiles for each of the 100 galaxies
which comprise the
ACS survey. In addition, we will examine their color profiles and
isophotal structure, search for the presence of embedded stellar disks, 
and examine their dust properties. In terms of
sample size, this survey exceeds those of Lauer et al. (1995) and
Rest et al. (2001), and is the first systematic study of the
inner regions of early-type galaxies to include color information.

\subsection{The Extragalactic Distance Scale}
\label{sec:distance}

Virgo is a complex, irregular cluster with a diameter of $\approx~10^{\circ}$.
The spatial distribution of known or suspected members, when combined with
radial velocity measurements for the brightest galaxies, reveals that
the cluster is, in fact, an aggregate of at least three distinct
subsystems associated with M87, M49 and M86 (VCC1316, VCC1226 and
VCC881, respectively). The structure of the cluster along the line of
sight remains a matter of debate, with estimates of the back-to-front
depth ranging from $\pm$6--8~Mpc (Young \& Currie 1995; 
Yasuda, Fukugita \& Okamura 1997) to $\pm$2 (Neilsen \& Tsvetanov 2000).
In principle, if accurate distances for a large sample of early-type cluster
members were available, it would be possible not only to measure the 
cluster depth, but to map out its full three-dimensional structure. 
Such information would be useful in refining the calibration of various
far-field Pop II distance indicators ($e.g.$, Tully-Fisher, fundamental plane), 
aiding in the interpretation of the cluster velocity field and mass 
modeling, 
and characterizing the intrinsic dispersion of those distance indicators
which have been applied to multiple Virgo members. 

Of the possible distance indicators, the method of surface brightness
fluctuations (SBF) seems best suited to the task (Tonry \& Schneider 1988;
Blakeslee, Ajhar \& Tonry 1999).
This method, in which distances are estimated from the ratio of the second
and first moments of their stellar luminosity functions, is both efficient
and accurate (with random errors of $\lesssim$ 1.5~Mpc at the distance of
Virgo; Tonry et al. 2001). Equally
important, it well suited to the measurement of distances
for early-type galaxies, including dwarfs. This latter issue is important
since the early-type dwarf galaxies in Virgo are the only class of galaxies
which are available in large numbers throughout the cluster, including its
core (Binggeli et al. 1985).

SBF distances (derived from $V$ and $I$ images) have now been measured for 
roughly two dozen giant ellipticals in Virgo (Neilsen \& Tsvetanov 2000;
Tonry et al. 2001). Recently, $B$ and $R$ images have been been used to measure
SBF distances for a sample of 16 dwarf ellipticals in Virgo (Jerjen, Binggeli \&
Barazza 2004). Despite the small samples, these studies make it clear that Virgo 
exhibits significant structure along the line of sight. For example, working
from the SBF catalog of Tonry et~al. (2001), West \& Blakeslee (2000) were
able to show that the brightest Virgo ellipticals fall along an axis which
passes roughly through M87 and M84: the cluster's so-called ``principal axis".
In a similar vein, Jerjen et al.
(2004) find evidence from their sample of dwarf ellipticals for two subclumps 
at distances of 15.8 and 18.5~Mpc, associated with M87 and M86,
respectively.

A key science goal of the ACS Virgo Cluster Survey
is the measurement of SBF distances for a large sample of Virgo
galaxies. The rationale behind the choice of filters for this
component of the survey is given in \S\ref{sec:filters}. Technical details on
the calibration of the $z$-band SBF method, the procedures used
in measuring fluctuation magnitudes from ACS images, and the scientific 
results on the extragalactic distance scale ($e.g.$, Virgo's three-dimensional
structure, a galaxy-by-galaxy comparison SBF and GCLF distances, etc.)
will be presented in future papers in this series.

\section{Sample Selection: Membership, Spatial Distribution and Luminosities}
\label{sec:properties}

The most complete and homogeneous survey of galaxies in the Virgo Cluster
remains that of Binggeli et~al. (1987) who used wide-field
photographic blue plates to identify and study galaxies within $\approx$
6$^{\circ}$ of the cluster center (defined by M87 = VCC1316).
Their Virgo Cluster Catalog (VCC) 
consists of 2096 galaxies within this $\approx$ 140 deg$^{2}$ region.
Membership in the cluster was established through a combination of morphological
classifications ($i.e.,$ members were identified on the basis of the 
relation between apparent magnitude and surface brightness) and radial
velocity measurements\footnote{For some late-type galaxies, cluster
membership could also be established through the resolution of star
forming knots.} (taken principally from the Virgo Cluster redshift
survey; Huchra 1985).
In addition to membership classifications and radial velocities,
Binggeli et al.  (1987) give coordinates, total magnitudes ($B_T$),
and morphologies for each of their VCC galaxies.

A total of 1277 VCC galaxies were considered by Binggeli et~al. (1987) to
be members of Virgo; an additional 574 galaxies were listed as possible 
cluster members (1851 galaxies in total). At the time of its preparation,
the VCC contained radial velocities for 403 cluster members. Of these, 352
galaxies have $B_T < 16$, which we take as the faint-end cutoff of
the ACS Virgo Cluster Survey. Early-type galaxies were selected from this
sample using the VCC morphological classifications of Binggeli \& Sandage
(1984). Specifically, program galaxies were required to have morphological
types of E, S0, dE, dE,N, dS0 or dS0,N. This leaves a total of 163 galaxies.
 
Due to the limited number of orbits awarded by the time allocation committee,
a subset of 100 galaxies were then selected for observation
with ACS. The 63 galaxies which were discarded include 13 dwarfs that 
were previously observed with WFPC2 as part of programs G0-5999, GO-6352
and GO-8600. The remaining galaxies were omitted for a variety of
reasons, such as uncertain morphologies, the lack of a clearly visible
bulge component, the presence of strong dust lanes, or signs of strong
tidal interactions.

\begin{figure}
\plotone{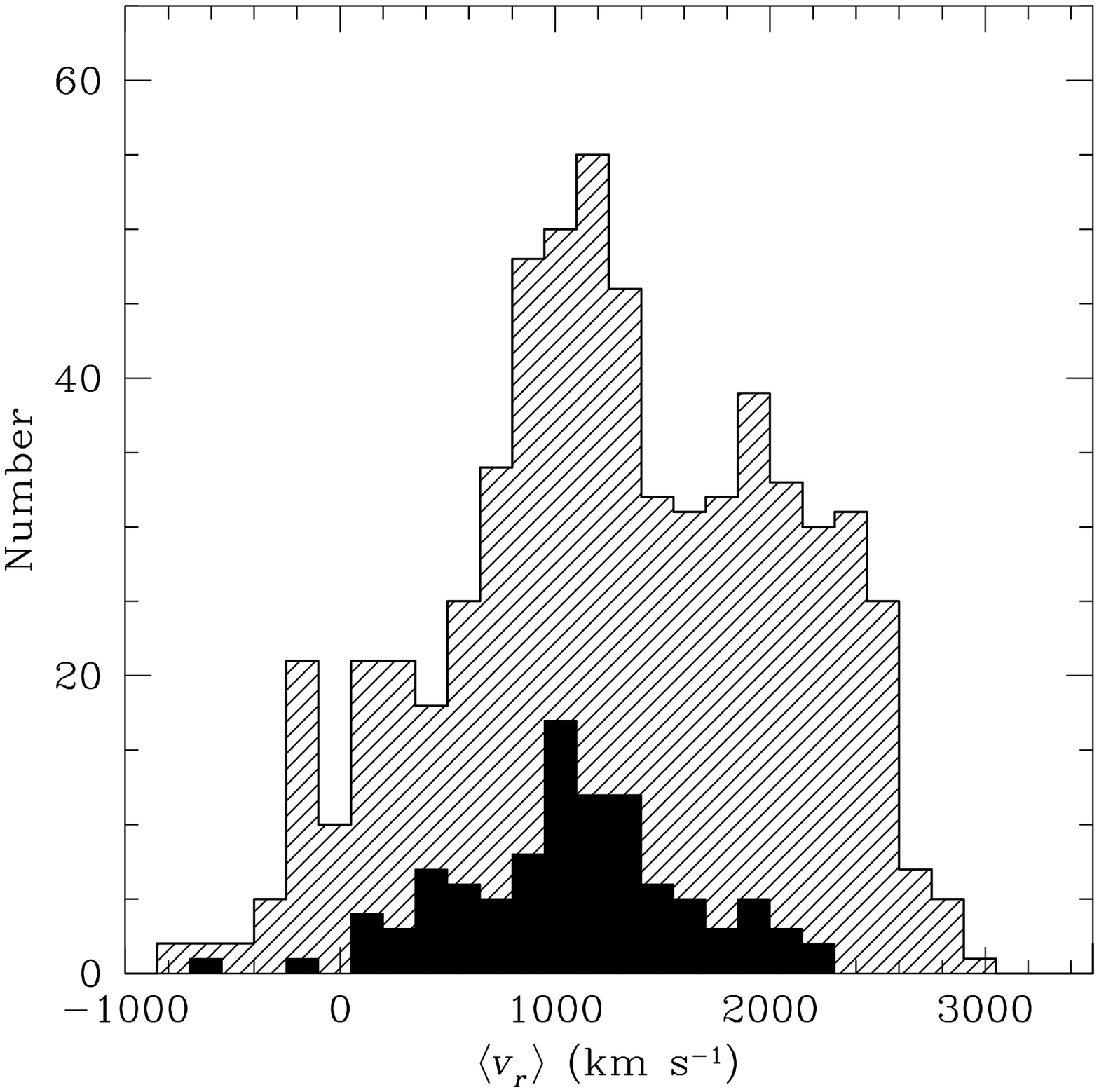}
\caption{Histogram of radial velocities for 897 galaxies, of all morphological types,
classified by
Binggeli $et~al.$ (1987) as members or possible members of the Virgo Cluster
{\it and} having measured radial velocities according to the NASA Extragalactic
Database (upper, hatched histogram). The filled lower histogram shows the
radial velocity distribution of ACS Virgo Cluster Survey program galaxies.
\label{fig01}}
\end{figure}

In the time since the VCC was compiled, Virgo has been the subject
of numerous spectroscopic surveys. As a result, radial velocities are now
available for many additional galaxies ($e.g.,$ Binggeli,
Popescu \& Tammann 1993; Drinkwater et al. 1996). According to NED,
897 of the 1851 galaxies originally classified by Binggeli
et al.  (1987) as certain or possible members of Virgo now have published
radial velocities. The upper histogram in Figure~\ref{fig01} shows the
distribution of radial velocities for these 897 galaxies, based on the
latest NED data. For comparison, the filled histogram shows the distribution
of radial velocities for ACS Virgo Cluster Survey galaxies. It is
clear that the velocity distribution for the ACSVCS program galaxies
closely resembles that of the entire Virgo sample.

\begin{figure}
\plotone{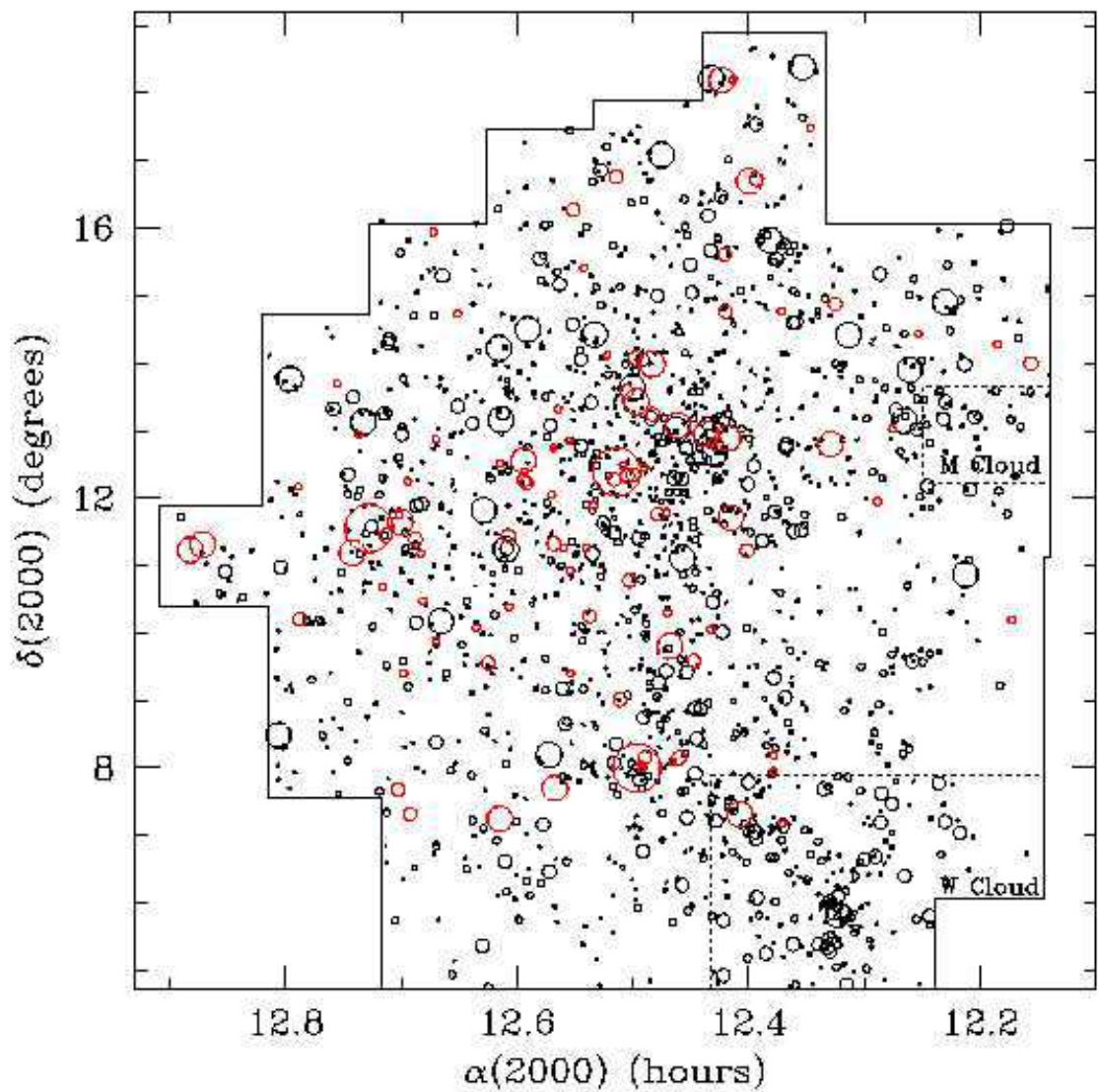}
\caption{Distribution of VCC galaxies on the plane of the sky, adapted from
Binggeli $et~al.$ (1987). The symbol size is proportional to blue luminosity.
This figure contains a total of 1726 galaxies, with no restriction on morphological
type, that are classified as
members or possible members of the Virgo Cluster and have declinations
greater than $\delta_{\rm B1950} \simeq 5^{\circ}$ (meaning that they are
not associated with the Southern Extension of Virgo). The M and W Clouds as
defined by Sandage $et~al.$ (1985) are shown by the dotted regions. Red
symbols denote the full sample of 100 early-type galaxies from the ACS
Virgo Cluster Survey.
\label{fig02}}
\end{figure}

The irregular structure of the cluster is apparent in
Figure~\ref{fig02}, which shows the distribution of 1726 VCC members or
possible members on the sky (black circles). This is somewhat fewer than
the 1851 members or possible members listed in the VCC since we show
only those galaxies that have $\delta_{\rm B1950} \ge 5^{\circ}$, meaning
that they are not associated with the Southern Extension of Virgo
(Sandage et al. 1985). For comparison, the red symbols show the 100
galaxies from the ACS Virgo Cluster Survey, which fall in the
declination range $7\fdg2 \le \delta_{\rm B1950} \le 18\fdg2$.

Sandage et al. (1985) examined the luminosity function of Virgo Cluster 
galaxies in detail, including its dependence on morphology. Numerous 
subsequent studies have used more sensitive photographic plates
($e.g.$, Impey, Bothun \& Malin 1988; Phillipps et al. 1998) or wide-field CCD
mosaic cameras  (Trentham \& Hodgkin 2002; Sabatini et al. 2003) to 
explore the faint-end behavior of the Virgo luminosity function and to search
for bright galaxies with compact sizes or extreme surface brightnesses
which might have gone undetected by Sandage et al. (1985).
While the precise form of the faint-end of the Virgo luminosity function 
remains a matter of debate, the various surveys generally agree on a
high level of completeness of the VCC for $B_T \lesssim 18$, the
completeness limit estimated by Binggeli et al. (1987).

\begin{figure}
\plotone{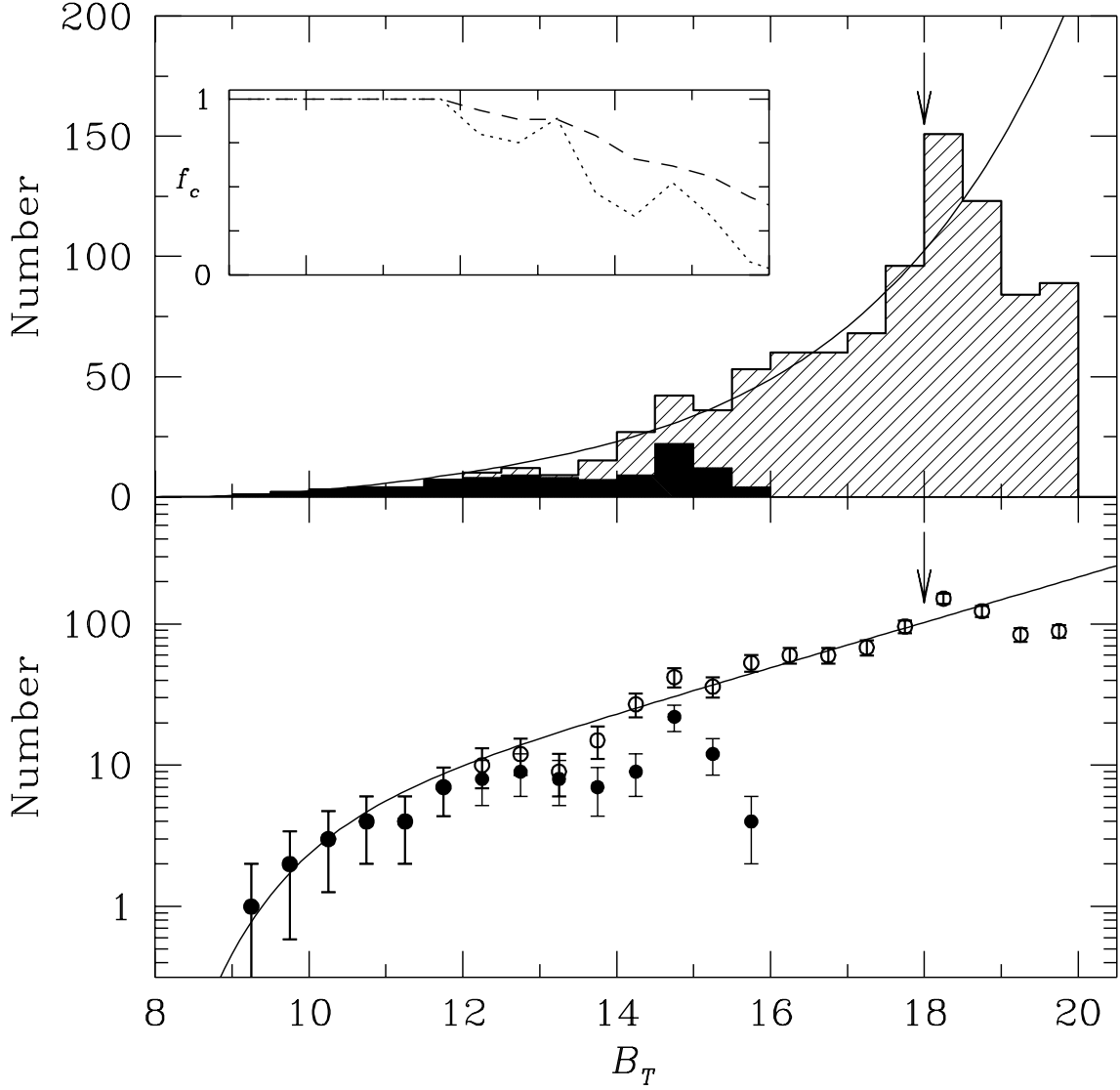}
\caption{{\it (Upper Panel)} Luminosity function of 956 early-type galaxies 
($T \le 15$) that are classified by Binggeli $et~al.$ (1987) as members of the Virgo Cluster 
(upper, hatched histogram). The arrow shows the VCC completeness limit, while
the solid curve shows the best-fit Schechter function for E+S0+dE+dS0 
galaxies from Sandage $et~al.$ (1985). The filled lower histogram shows the 
luminosity function of the 100 early-type galaxies in the
ACS Virgo Cluster Survey. 
{\it (Inset to Upper Panel)} Differential (dotted curve) and cumulative 
(dashed curve) completeness fractions for the ACS Virgo Cluster Survey
over the range $9 \le B_T \le 16$.
{\it (Lower Panel)} Same as above, except in logarithmic form. The open
circles show the luminosity function of 956 early-type members of the 
Virgo Cluster according to Sandage $et~al.$ (1985). The arrow 
shows the VCC completeness limit of the original VCC, while the 
filled circles show the luminosity
function of galaxies in the ACS Virgo Cluster Survey. The solid
curve is the same as that shown in the upper panel.
\label{fig03}}
\end{figure}

The upper panel of Figure~\ref{fig03} shows the luminosity function of
956 early-type galaxies judged by Binggeli et al. (1987) to be 
members of Virgo. The lower panel of this figure shows
the same luminosity function in logarithmic form. The filled 
histogram (upper panel) and filled circles (lower panel) show the 
corresponding luminosity functions for the ACS Virgo Cluster
Survey program galaxies, which have $9.31 \le B_T \le 15.97$ 
(corresponding to a factor of $\sim$~450 in luminosity). Note
that these galaxies are all considerably brighter than
the VCC completeness limit of $B_T \approx 18$ (indicated by the
arrows in Figure~\ref{fig03}). The inset in the upper panel of 
this figure shows the completeness of the ACS Virgo Cluster Survey
as a function of magnitude --- the dotted curve shows the completeness 
in 0.5 mag bins, while the dashed curve shows the variation in 
cumulative completeness. The final sample of galaxies 
constitutes a complete sample of early-type members of the Virgo Cluster
with $B_T < 12$. Brighter than $B_T \simeq 16$, the survey includes
44\% of the early-type members of Virgo.

General properties for the program galaxies are
given in Table~\ref{tab1}. From left to right, this table records
the magnitude ranking from 1$-$100 (which also serves as the 
identification number for each galaxy in the survey),
VCC number, right ascension and declination, blue magnitude, 
heliocentric radial velocity from NED, morphological type
from NED, and alternative names in the Messier, NGC, UGC or IC
catalogs. Table~\ref{tab2} gives the breakdown of this sample in
terms of morphology; roughly two thirds of the galaxies in the
survey (65 objects) are elliptical (E) or lenticular (S0) galaxies. 
The remaining third of the sample (35 objects) are dwarfs of
type dE, dE,N, or dS0. 

\section{Observational Strategy}
\label{sec:strategy}

The observing strategy adopted for the ACS Virgo Cluster Survey was chosen
with the scientific goals from \S\ref{sec:motivation} in mind.
Critical issues in planning the observing program include the
selection of program galaxies (discussed in \S\ref{sec:properties}),
and the choice of filters, exposure times and field orientation.
These issues are discussed in detail below. 

\subsection{Choice of Filters}
\label{sec:filters}

Two filters are required since color information is an essential
ingredient in all three science programs. That is to say,
multi-wavelength imaging is needed to select GC candidates and 
to estimate their metallicities, measure color profiles for
the galaxies and examine their dust properties,
and to quantify the stellar population dependence of the
SBF magnitudes. The choice of filters is driven by
the desire for high-throughput and the need for 
exceptional metallicity and age sensitivity when modeling
the underlying stellar populations.

\begin{figure}
\plotone{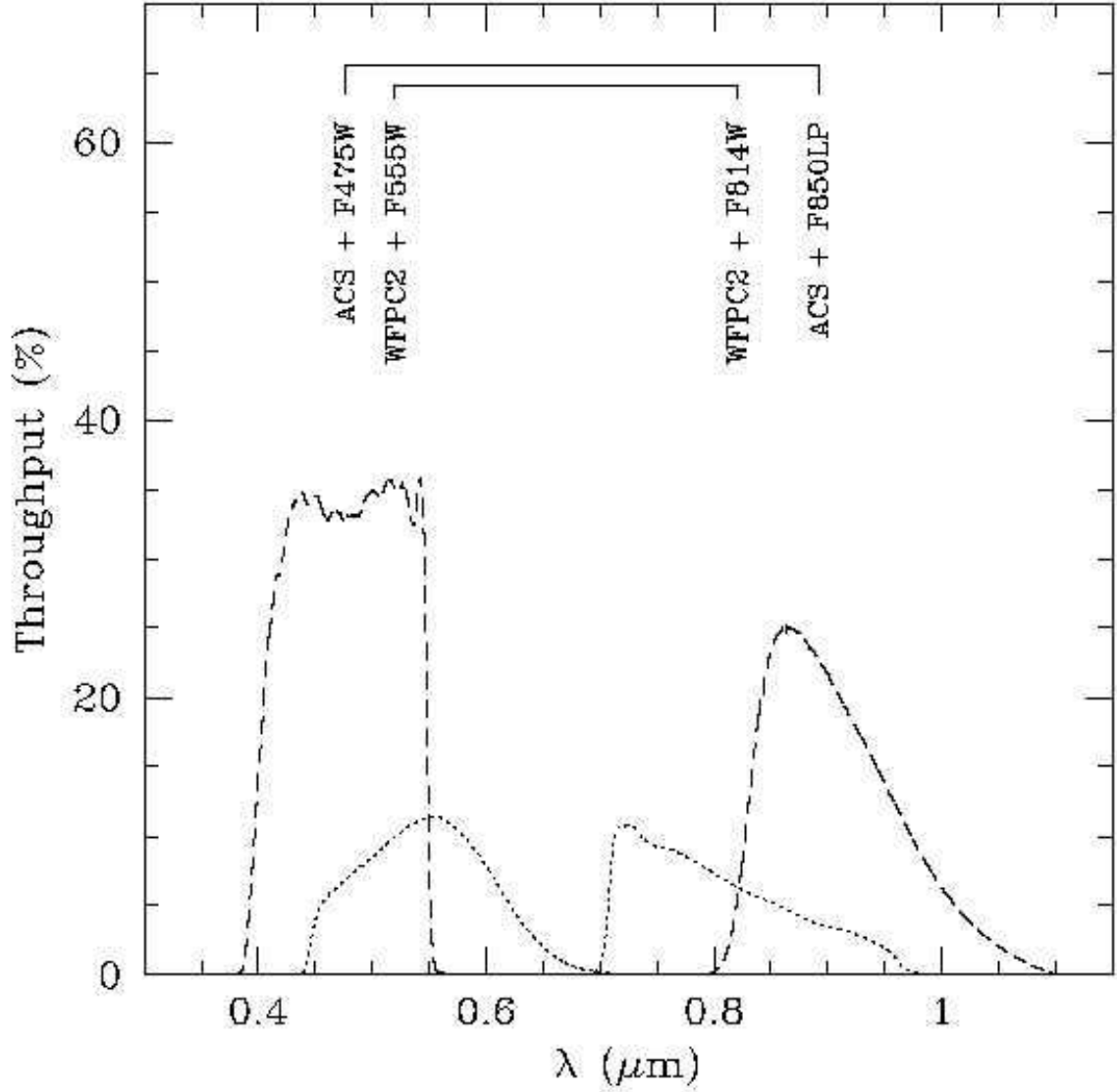}
\caption{Total system throughput for ACS/WFC in the F475W
and F850LP bandpasses (dashed curves). For comparison, the dotted
curves show the WFPC2 throughput for F555W and F814W, the HST camera/filter
combinations used most frequently in the measurement of SBF distances and in
studies of extragalactic globular cluster systems.
\label{fig04}}
\end{figure}

The F475W and F850LP filters, which offer a combination of high
throughput and excellent age and metallicity sensitivity, are 
equivalent to the $g$ and $z$ filters in the Sloan Digitized Sky
Survey system. Transmission curves
for the two filters are shown in Figure~\ref{fig04}, which
also includes a comparison to the F555W ($\approx V$) and
F814W ($\approx I$) filters on WFPC2. This latter filter/camera
combination is the most commonly used in previous HST studies of
extragalactic GC systems and SBF measurements ($e.g.$, Gebhardt \&
Kissler-Patig 1999; Kundu \& Whitmore 2001; Larsen et al. 2001;
Ajhar et~al. 1997; Neilsen \& Tsvetanov 2000). The 
throughput advantage of F475W and F850LP on the ACS
relative to F555W and F814W on WFPC2 is a factor of
2--4, with an increase in color baseline 
of $\approx$~50\%.

\begin{figure}
\plotone{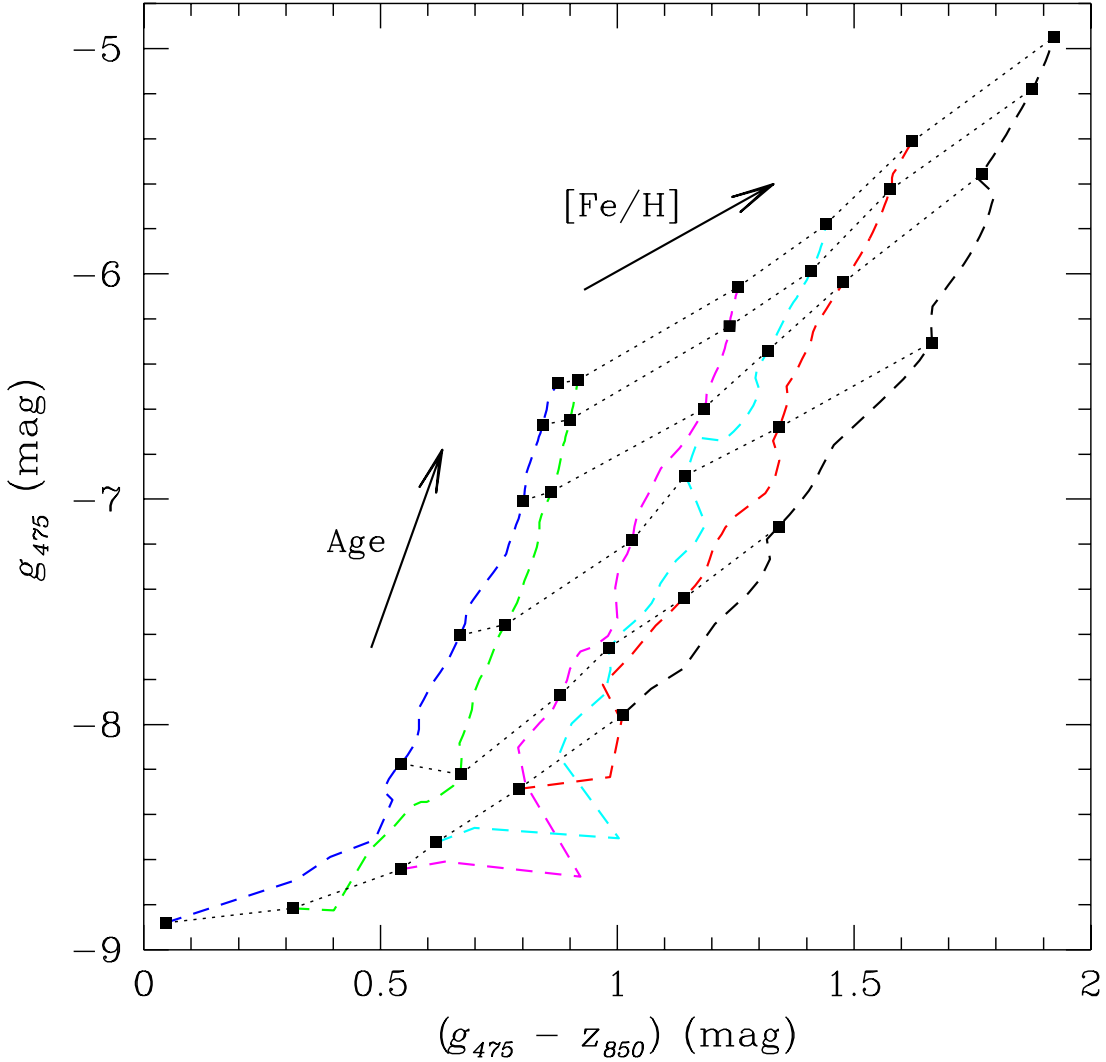}
\caption{Simple stellar populations in the
$g_{\rm 475}$ versus ($g_{\rm 475}-z_{\rm 850}$) plane calculated
using the population synthesis models of Bruzual \& Charlot (2003). The dashed grid
lines represent isometallicity tracks of [Fe/H] = --2.25 (blue), --1.65 (green), --0.64
(magenta), --0.33 (cyan), +0.09 (red) and +0.56 (black). The dotted grid lines show
isochrones of age T = 1, 2, 4, 8, 12 and 15 Gyr. The $g_{\rm 475}$
magnitudes have been scaled to a total mass of ${\cal M} = 2.4\times10^5{\cal M}_{\odot}$,
the approximate mean mass of globular clusters in the Milky Way.
\label{fig05}}
\end{figure}

\begin{figure}
\plotone{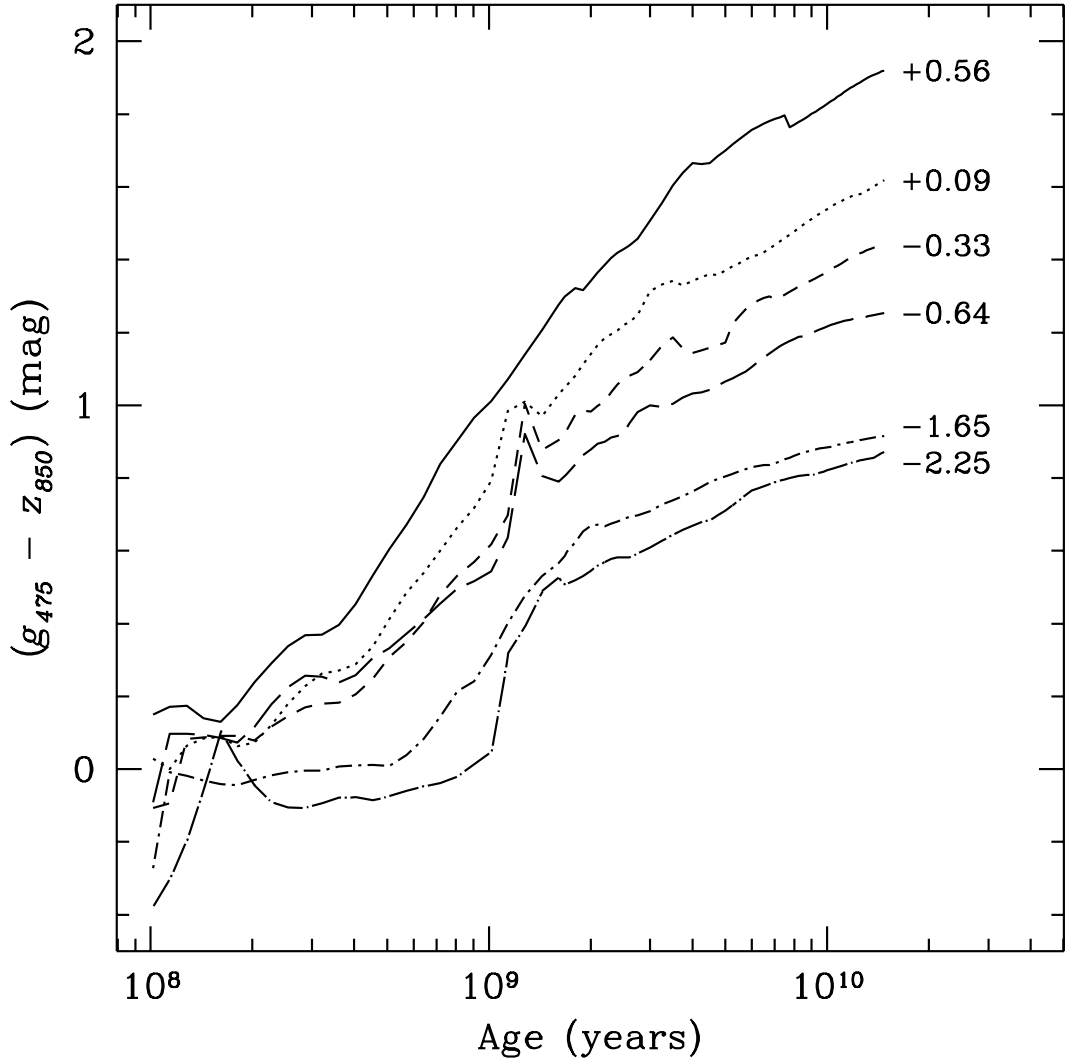}
\caption{Time evolution of ($g_{\rm 475}-z_{\rm 850}$) color for simple
stellar populations calculated using the population synthesis models of Bruzual \& Charlot
(2003). The curves show isometallicity tracks of [Fe/H] = --2.25, --1.65,
--0.64, --0.33, +0.09 and +0.56~dex.
\label{fig06}}
\end{figure}

Figures~\ref{fig05} and \ref{fig06} show the predicted age and
metallicity sensitivity of this filter combination based on
the population synthesis models of Bruzual \& Charlot (2003),
assuming a Salpeter IMF.
The magnitudes and colors shown here were determined using the
F475W and F850LP filter transmission curves of Sirianni et al.
(2004; private communication). Figure~\ref{fig05} shows the behavior,
in the $g_{\rm 475}$ versus ($g_{\rm 475}-z_{\rm 850}$) plane,
of simple stellar population models normalized to
a total mass of ${\cal M} = 2.4\times10^5{\cal M}_{\odot}$ (appropriate
for the mean mass of the Galactic GCs; McLaughlin
1999b). The time evolution in
($g_{\rm 475}-z_{\rm 850}$) color of these
model stellar populations over the age range 100 Myr $\le {\rm T} \le$ 15 Gyr is
shown in Figure~\ref{fig06}. If we approximate the behavior
by linear relations between age and color, then 
a least-squares fit to the model predictions in Figure~\ref{fig06} gives
$${\rm 0.57} \lesssim {\partial{(g_{\rm 475}-z_{\rm 850})} \over \partial{\rm \log{T}}} \lesssim {\rm 0.91~mag~dex}^{-1}$$
for metallicities between
$-2.25 \lesssim$ [Fe/H] $\lesssim +0.56$~dex.
The corresponding dependence in
($V-I$) color is $0.33 - 0.51$~mag~dex$^{-1}$.
For comparison, the metallicity sensitivity of the 
($g_{\rm 475}-z_{\rm 850}$) index at a constant
age of 10 Gyr is,
$${\partial{(g_{\rm 475}-z_{\rm 850})} \over \partial{\rm [Fe/H]}} \simeq {\rm 0.36~mag~dex}^{-1},$$
or twice that of ($V-I$), which has slope $\simeq$ 0.18~mag~dex$^{-1}$.

This filter combination is also advantageous when trying to determine 
the incidence of dust in the target galaxies.  Being bluer, dust
absorption in $g_{\rm 475}$ is 0.16 mag larger
than in $V$, assuming a standard reddening law (Cardelli, Clayton 
\& Mathis 1989). In addition, the longer wavelength baseline of the
$(g_{\rm 475}-z_{\rm 850})$ color, compared to the more commonly adopted
$(V-I)$ color, results in a 50\% increase in the contrast produced in the
color images by a given amount of dust obscuration.

\begin{figure}
\plotone{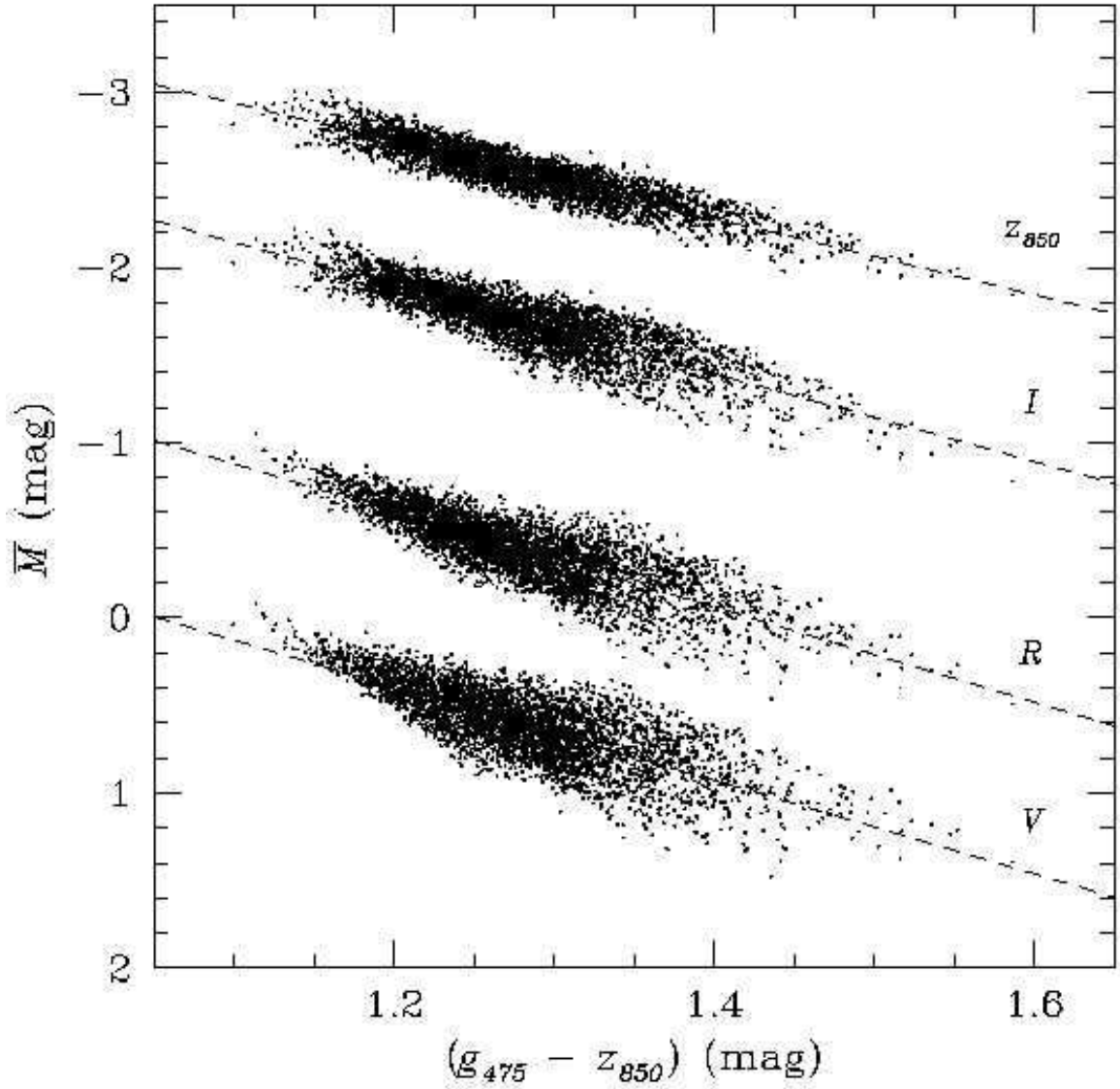}
\caption{Fluctuation magnitude, ${\overline M}$, in $V,~R,~I$ and $z_{\rm 850}$
calculated using the composite stellar population models of Blakeslee,
Vazdekis \& Ajhar (2001) and transformed to $z_{\rm 850}$ via
the relations of Fukugita et al. (1996).
\label{fig07}}
\end{figure}

Since a primary goal of the survey is the measurement of accurate
SBF distances, this has also driven the choice of filters. A number
of considerations suggest that, in general, the $z$ band
is a logical choice for SBF studies.  Because the SBF signal
in early-type galaxies is dominated by evolved stars, the 
fluctuations are quite red (having typical colors of
$\overline{V}$-$\overline{I}$ $\sim$ 2.4; Ajhar \& Tonry 1994). 
Figure~\ref{fig07} compares fluctuation magnitudes, $\overline{M}$, in 
$V$, $R$, $I$ and $z_{\rm 850}$ predicted by the models
of Blakeslee, Vazdekis \& Ajhar (2001). It is evident that the
SBF in $z_{\rm 850}$ is not only bright ($\approx 0.9$~mag
brighter than in the $I$-band) but well behaved, with an rms
scatter of only 0.066~mag (compared to 0.093~mag in the
$I$ band).\footnote{By this rationale, one might expect the
near-infrared to be even better for SBF. However, longward of 1$\mu$,
the effects of age and metallicity become non-degenerate on $\overline{M}$,
leading to increased scatter in the calibration (Blakeslee,
Vazdekis \& Ajhar 2001).}
In addition, the $z$ band is 20\% less affected by dust extinction
than the $I$-band, allowing the optical SBF technique to be pushed to
lower Galactic latitudes. A complete discussion of the 
implementation and calibration of the $z$-band SBF technique
will be presented in a future paper.

\subsection{Exposure Times and Limiting Magnitudes}
\label{sec:limiting}

A single orbit with HST was devoted to each galaxy. Target visibility
for Virgo is $\sim$ 3200 sec, and a variety of overheads ($e.g.$, guide
star acquisitions, filter changes and detector readouts) further limit
the usable exposure time. Exposure times in the two filters were
set by the requirement that the limiting magnitudes in
$g_{\rm 475}$ and $z_{\rm 850}$ reach roughly equal depths for
old, metal-rich stellar populations. For each galaxy, we obtained
a pair of exposures totaling 720 sec in F475W and 1120 sec in F850LP,
CR-SPLIT to aid in the rejection of cosmic rays.
An additional 90 sec exposure in F850LP was also
taken for each galaxy, in the event that the galaxy nucleus 
saturated the detector in the red bandpass.

\begin{figure}
\plotone{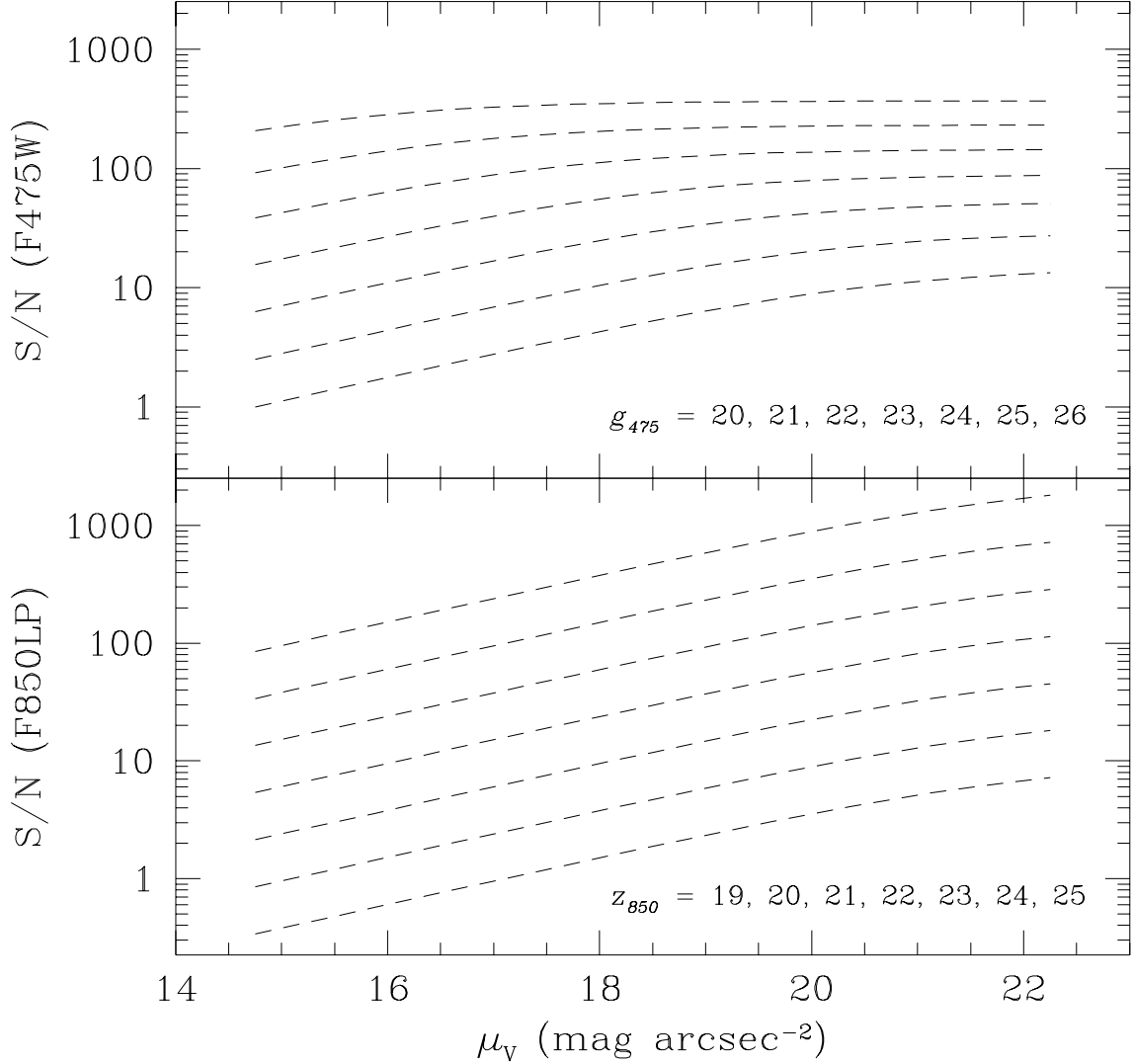}
\caption{{\it (Upper Panel)} Point-source signal-to-noise ratio, S/N, achieved in
a total exposure time of 2$\times$375 = 750 sec in the F475W
filter. The signal-to-noise ratio measured in a 3$\times$3 pixel aperture is
plotted as a function of the sky brightness, $\mu_V$.
From top to bottom, the seven curves correspond to point sources of
$g_{\rm 475}$ = 20, 21, 22, 23, 24, 25 and 26.
{\it (Lower Panel)} Same as above, except for a total exposure time of
2$\times$560 = 1120 sec in the F850LP filter.
From top to bottom, the seven curves correspond to point sources of
$z_{\rm 850}$ = 19, 20, 21, 22, 23, 24 and 25.
\label{fig08}}
\end{figure}

The limiting magnitude of these images is set not only by
the choice of filters and exposure times, but also by the 
background ``sky" level which varies both from galaxy to galaxy
and as a function of pixel position for individual galaxies.
Accurate results for the true completeness limits and 
photometric uncertainties require detailed artificial star
experiments; the results of such experiments will be presented in
a future paper. For the time being, we show in Figure~\ref{fig08}
the signal-to-noise ratio, S/N, expected for point sources as a
function of total sky brightness, $\mu_V$. Results for
the F475W and F850LP filters are shown in the
upper and lower panels, respectively. The dashed curves show
the S/N for exposure times of 750s in F475W and 1120s in F850LP.
At effective radii of 10$^{\prime\prime}$,
the surface brightness of the program galaxies is expected
to be $\mu_V \sim 19$~mag~arcsec$^{-2}$ or less, so that 
point-sources brighter than $g_{\rm 475} \sim 25.5$ or 
$z_{\rm 850} \sim 23.5$ should be detected with S/N $\gtrsim 10$.
In regions of low background, the detection limits are expected
to be $g_{\rm 475} \sim 26.1$ and
$z_{\rm 850} \sim 24.8$, meaning that the brightest
$\sim$ 95\% of the GCLF lies above the detection threshold.

An automated data reduction pipeline has been
written to align and combine the images, 
perform object detection, and measure object magnitudes, colors
and structural parameters. A complete description of this
pipeline is given in Jord\'an et al. (2004b; Paper II).

\subsection{Field Placement and Orientation}
\label{sec:field}

Coordinates for the nucleus of each galaxy were taken from NED.
While the precise centering of the galaxy within the ACS/WFC field 
is not crucial for the study of
the GC systems or the determination of fluctuation magnitudes,
it is critical for the measurement of nuclear surface brightness
profiles. For instance, ``break" radii for the
brightest Virgo galaxies can approach several arcseconds
(Byun et al. 1995), so it is important that the nuclei be placed
well away from the gap between the WFC1 and WFC2 detectors. This
is particularly important given the fact that uncertainties
in the NED coordinates can approach several arcseconds for a
number of the galaxies.
For each galaxy, the nucleus was therefore centered
on the WFC aperture, which has a reference pixel position at
(2096,~200) on the WFC1 detector. For the twelve brightest
galaxies, the nucleus was then
offset by 10$^{\prime\prime}$ away from the gap using
the POS TARG command, giving a distance of
$\approx 20^{\prime\prime}$ between the gap and nucleus. For
the remaining galaxies, a POS TARG offset of 5$^{\prime\prime}$
was used, giving a typical gap-nucleus distance
of $\approx 15^{\prime\prime}$.

\begin{figure}
\plotone{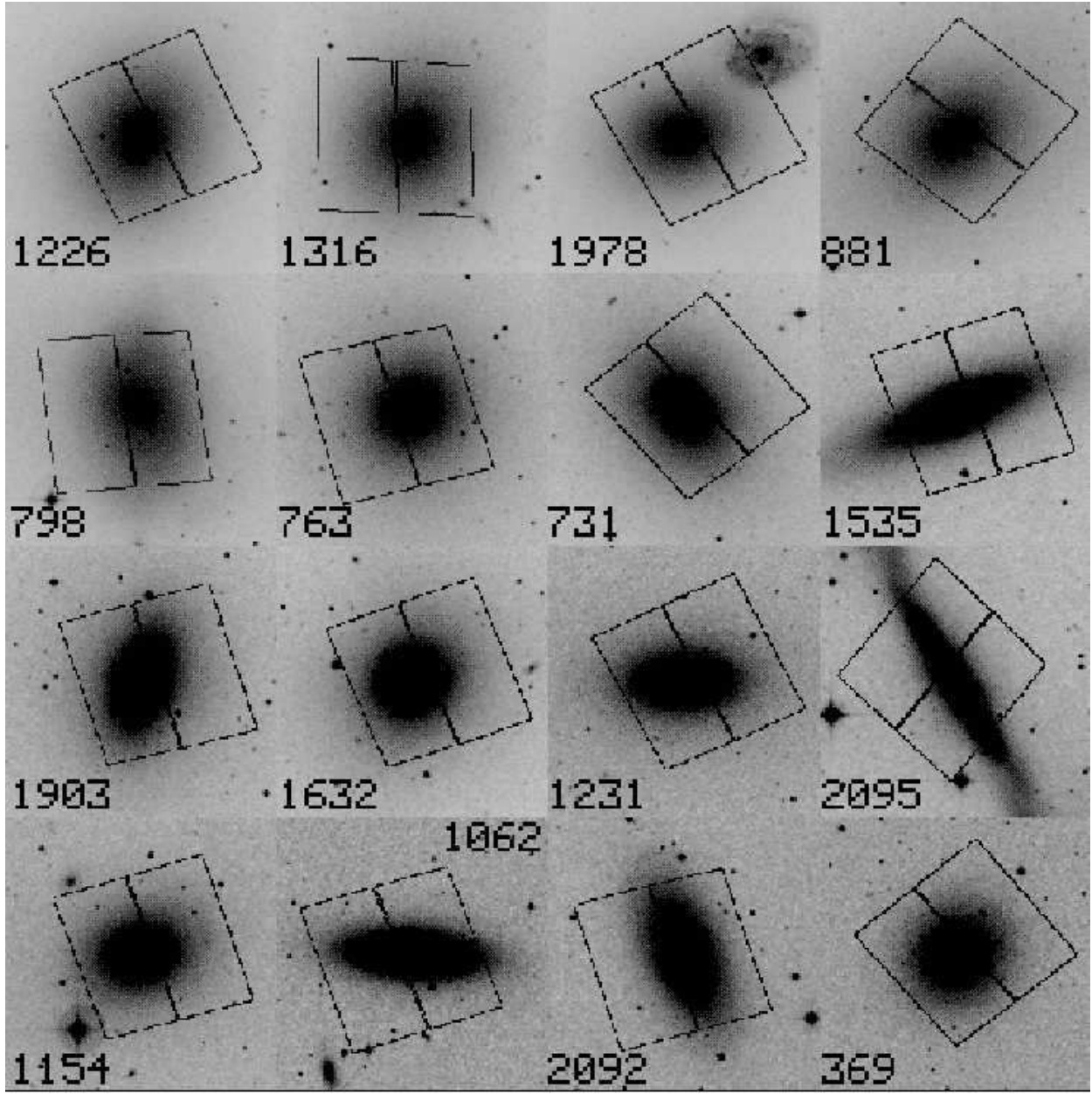}
\caption{Digitized Sky Survey images for ACS Virgo Cluster Survey galaxies \#1--16.
Images measure $6^{\prime}\times6^{\prime}$ and are oriented so that North is up and East 
is to the left. VCC identifications are labelled in each panel, while the overlay 
shows the location and orientation of the ACS/WFC field of view. 
\label{fig09}}
\end{figure}

\begin{figure}
\plotone{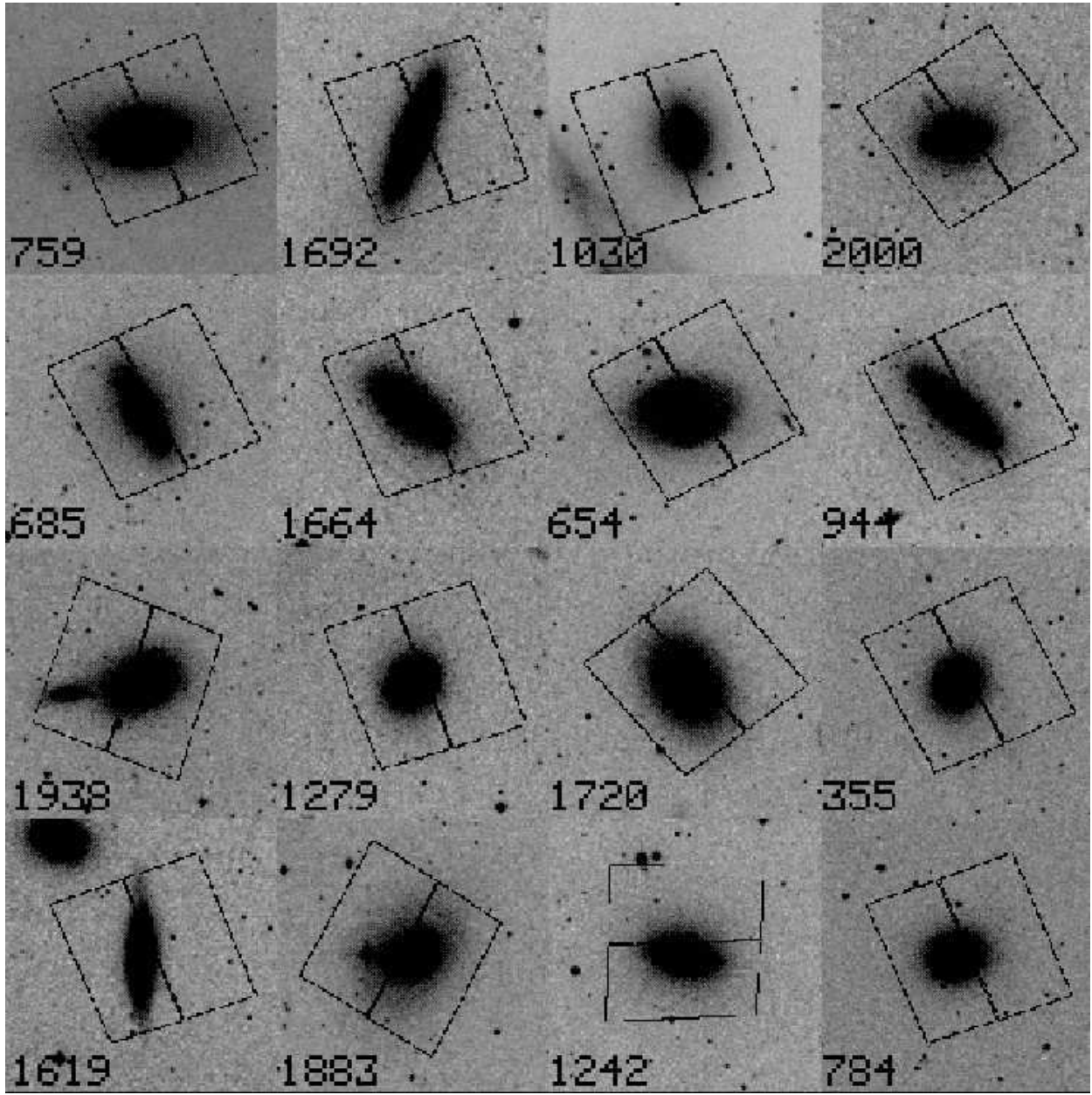}
\caption{Same as Figure~\ref{fig09}, except for galaxies \#17--32.
\label{fig10}}
\end{figure}

\begin{figure}
\plotone{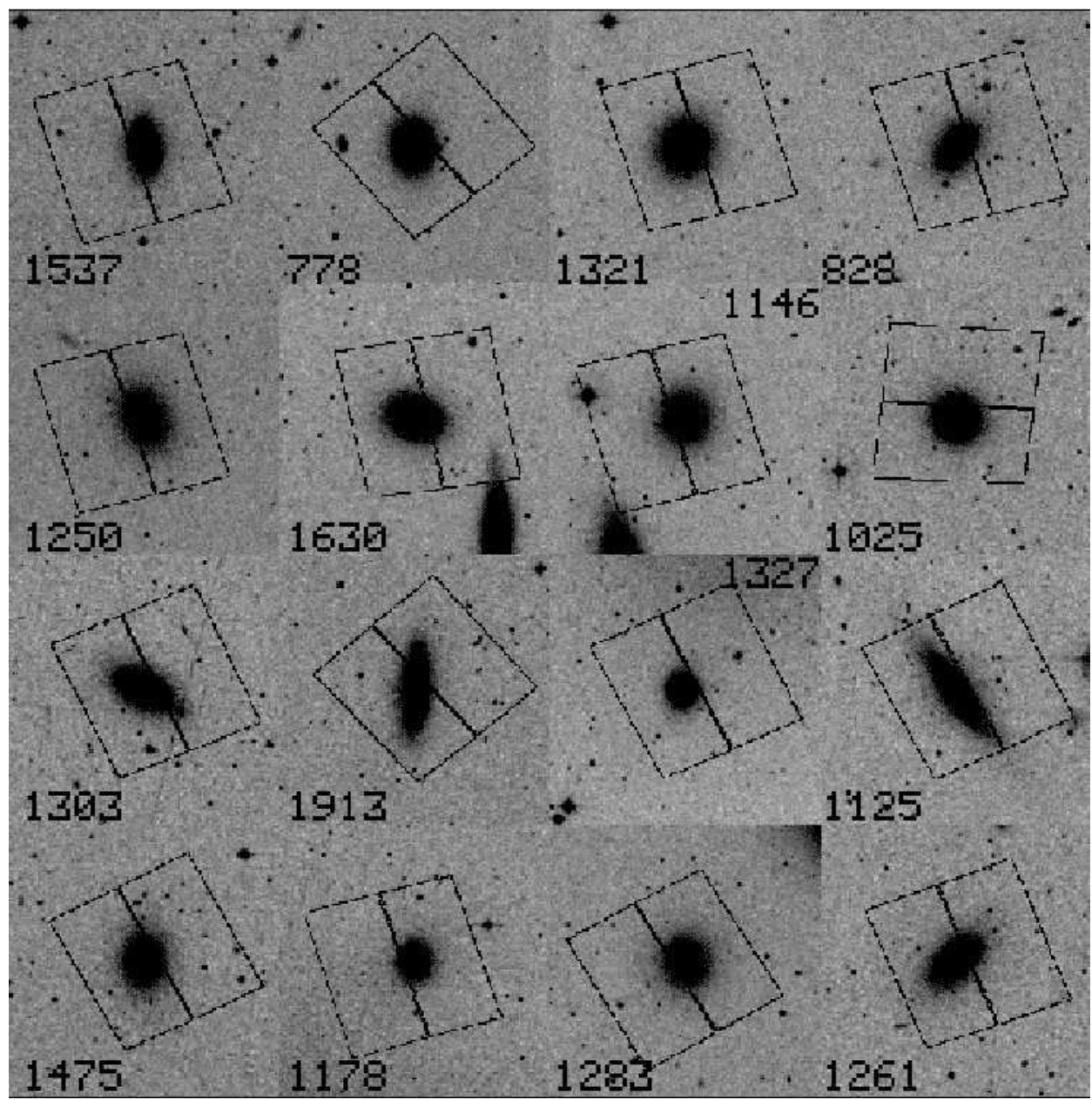}
\caption{Same as Figure~\ref{fig09}, except for galaxies \#33--48.
\label{fig11}}
\end{figure}

\begin{figure}
\plotone{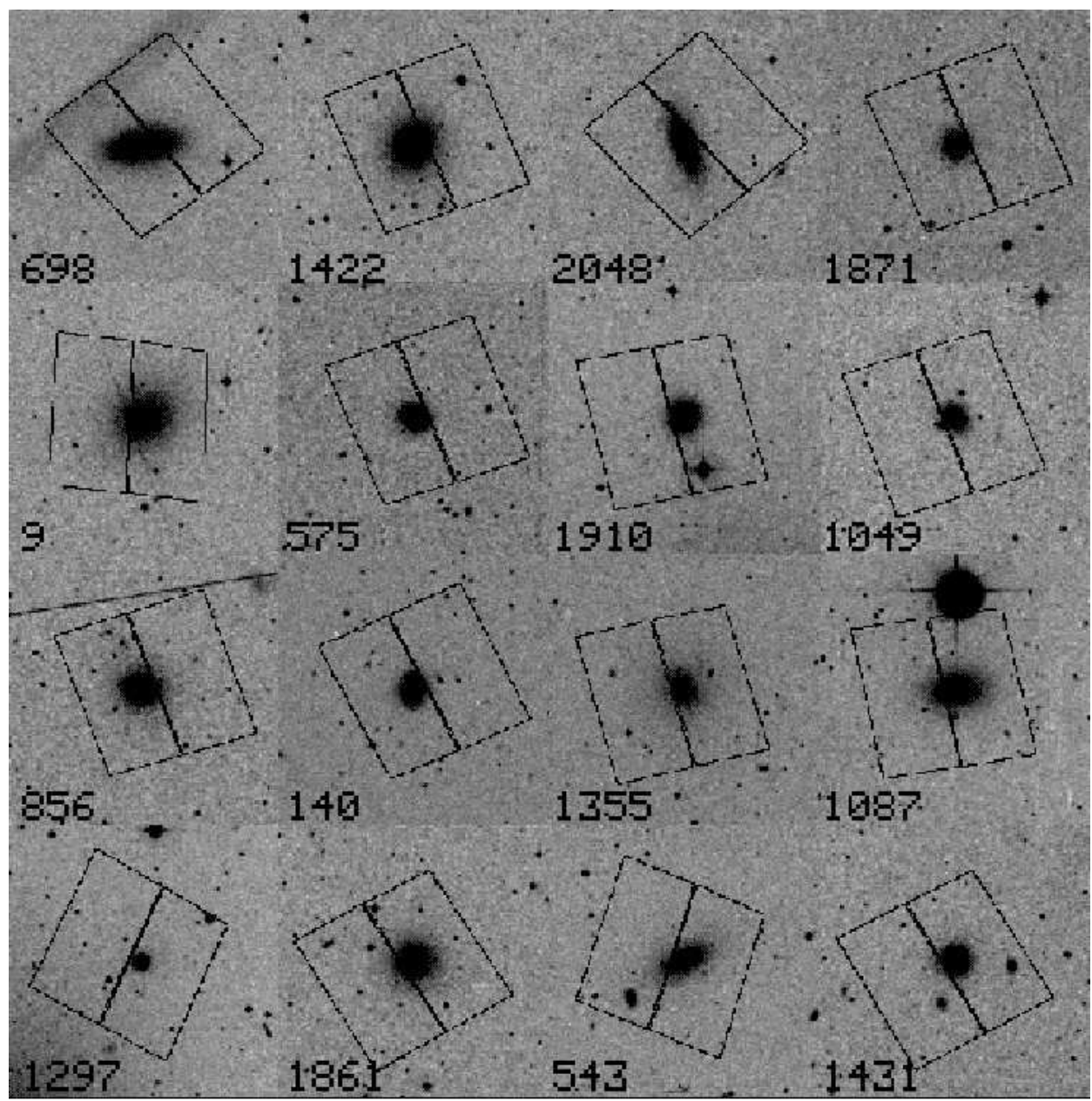}
\caption{Same as Figure~\ref{fig09}, except for galaxies \#49--64.
\label{fig12}}
\end{figure}

\begin{figure}
\plotone{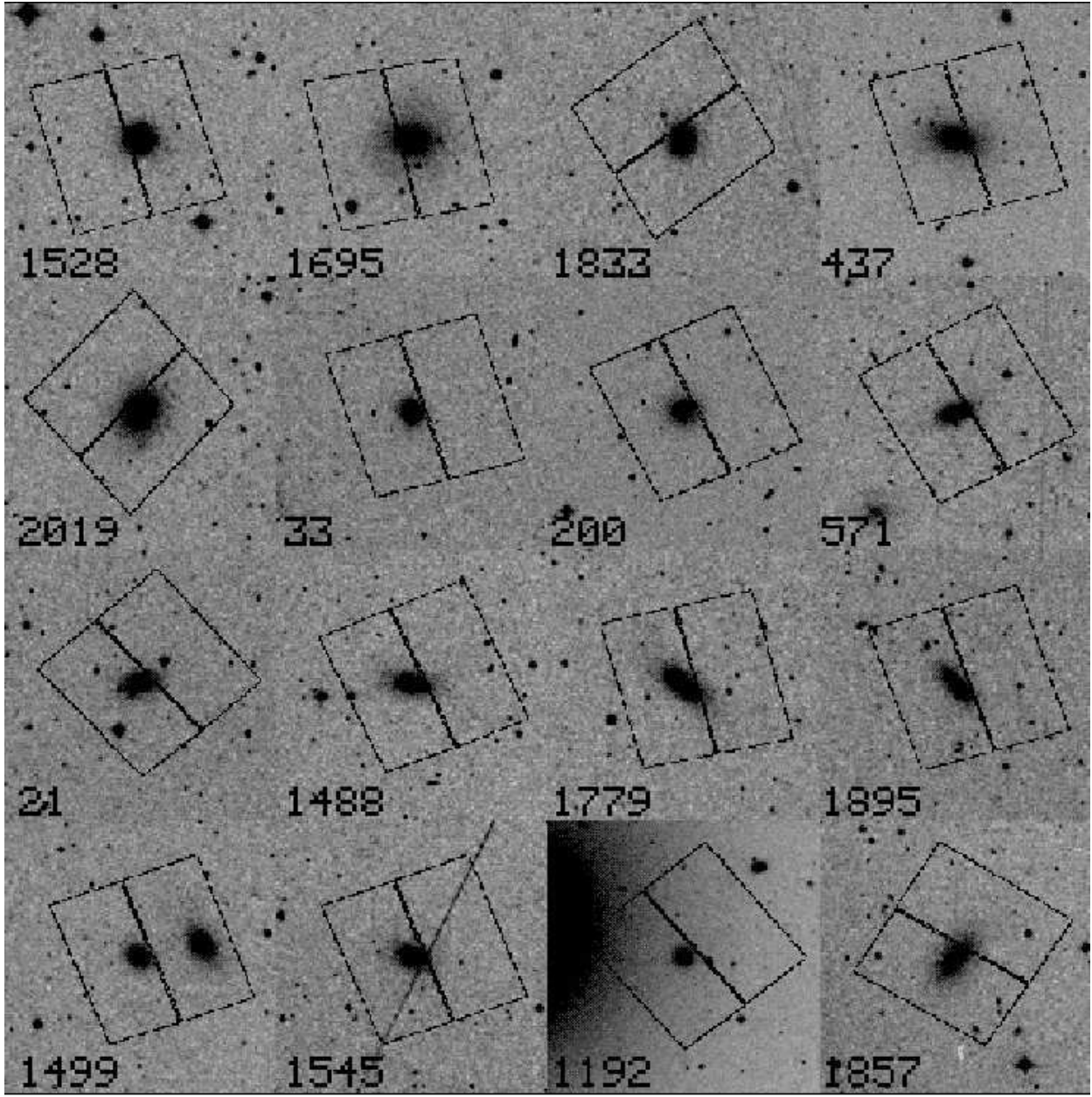}
\caption{Same as Figure~\ref{fig09}, except for galaxies \#65--80.
\label{fig13}}
\end{figure}

\begin{figure}
\plotone{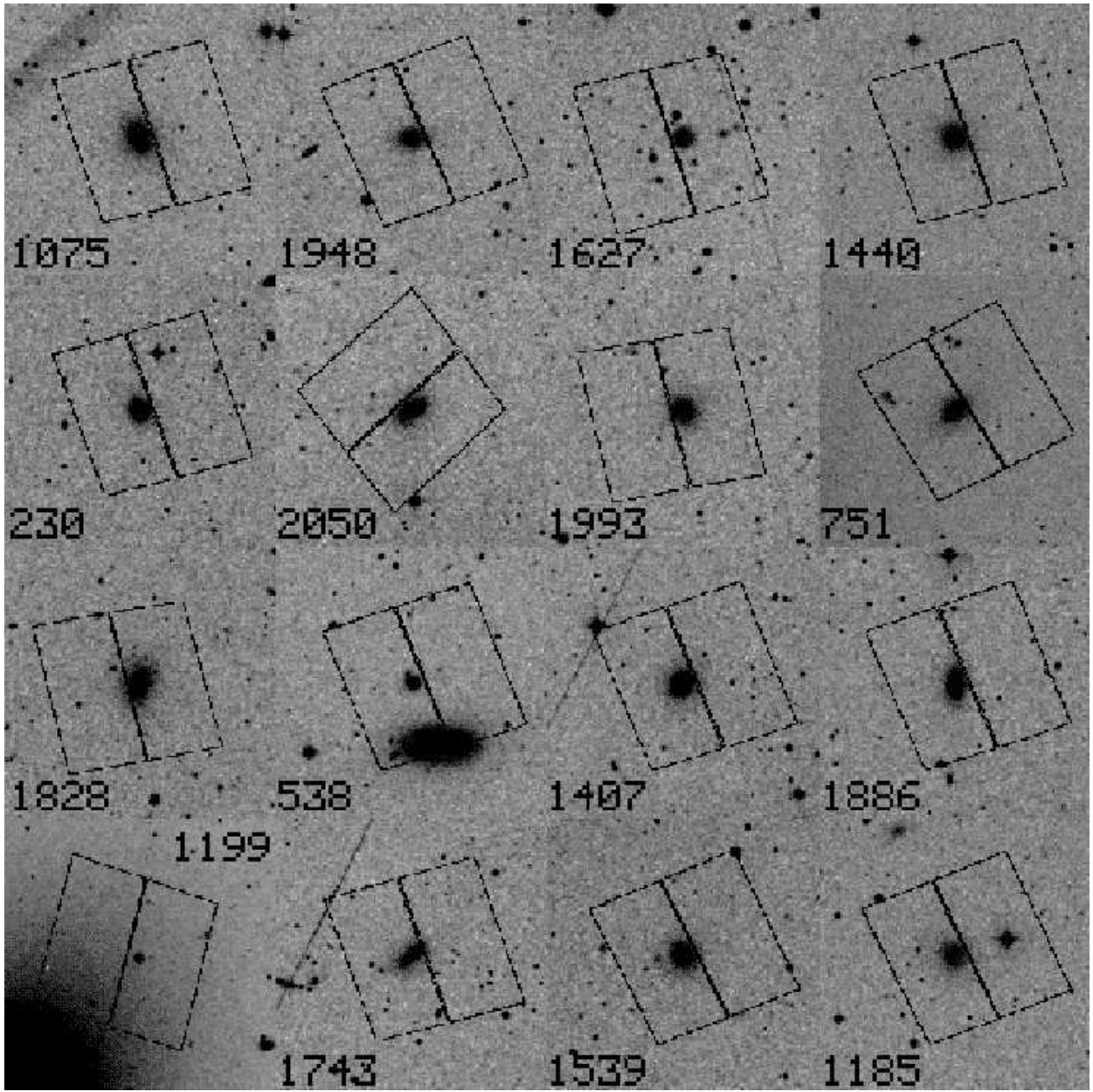}
\caption{Same as Figure~\ref{fig09}, except for galaxies \#81--96.
\label{fig14}}
\end{figure}

\begin{figure}
\plotone{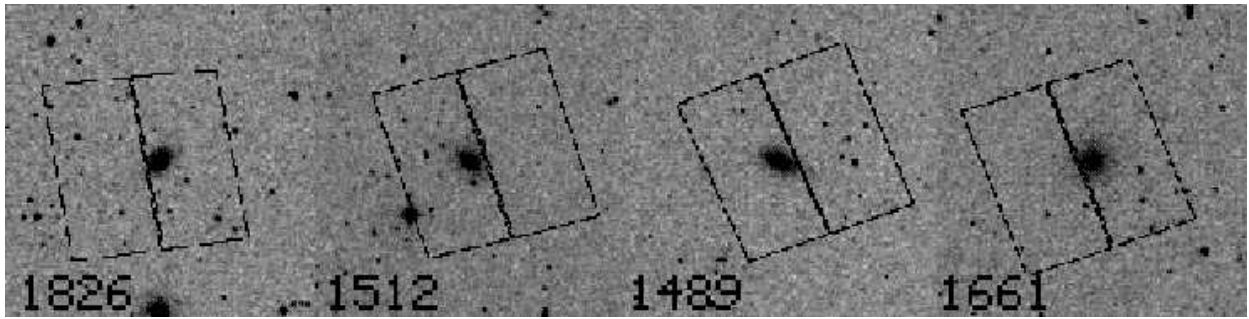}
\caption{Same as Figure~\ref{fig09}, except for galaxies \#97--100.
\label{fig15}}
\end{figure}

\begin{figure}
\plotone{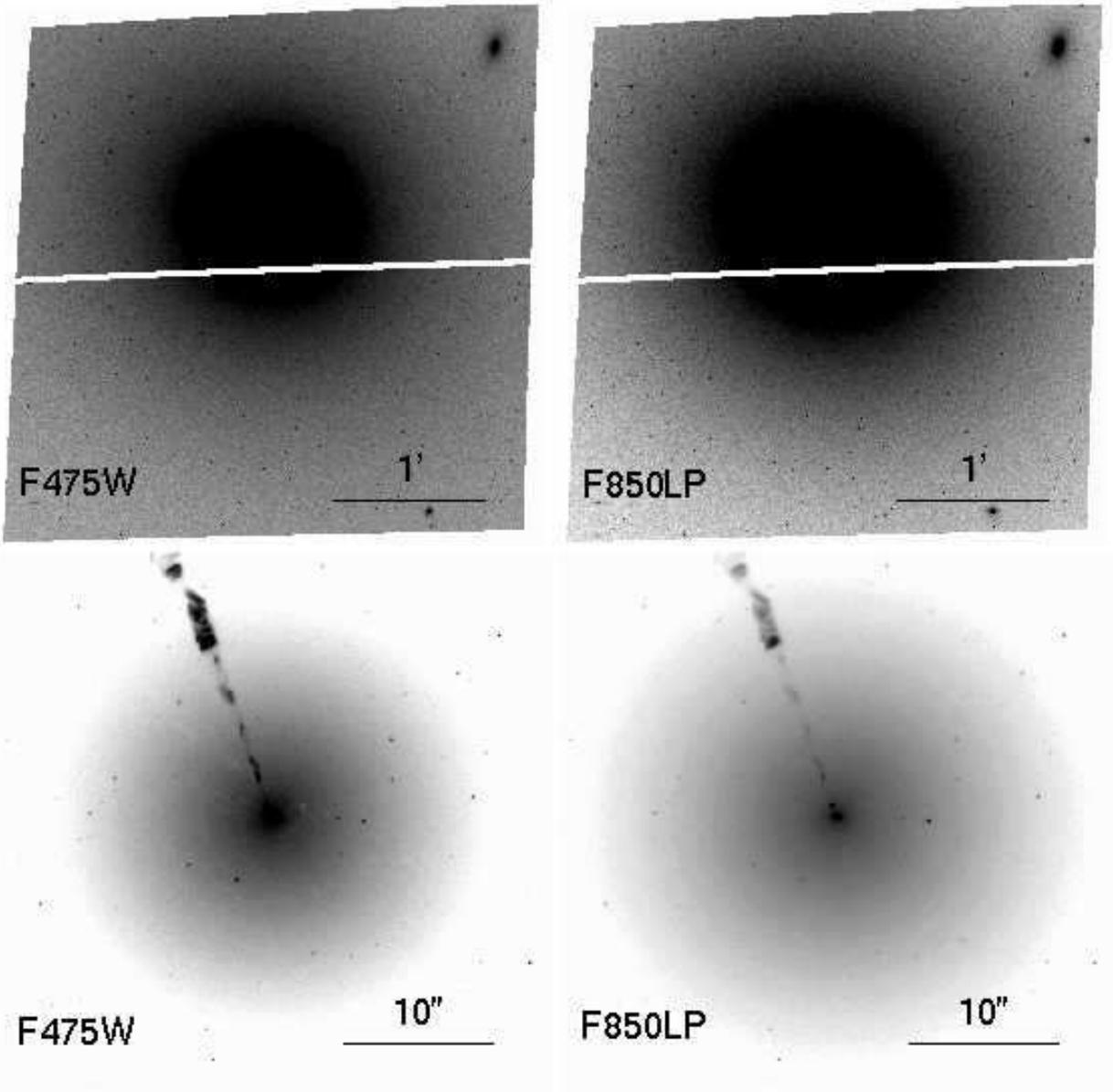}
\caption{F475W and F850LP images for VCC1316 (M87).
The upper panels show the full ACS/WFC images, while the bottom panels show 
$36^{\prime\prime}\times36^{\prime\prime}$ regions centered on the galaxy's nucleus.
\label{fig16}}
\end{figure}

\begin{figure}
\plotone{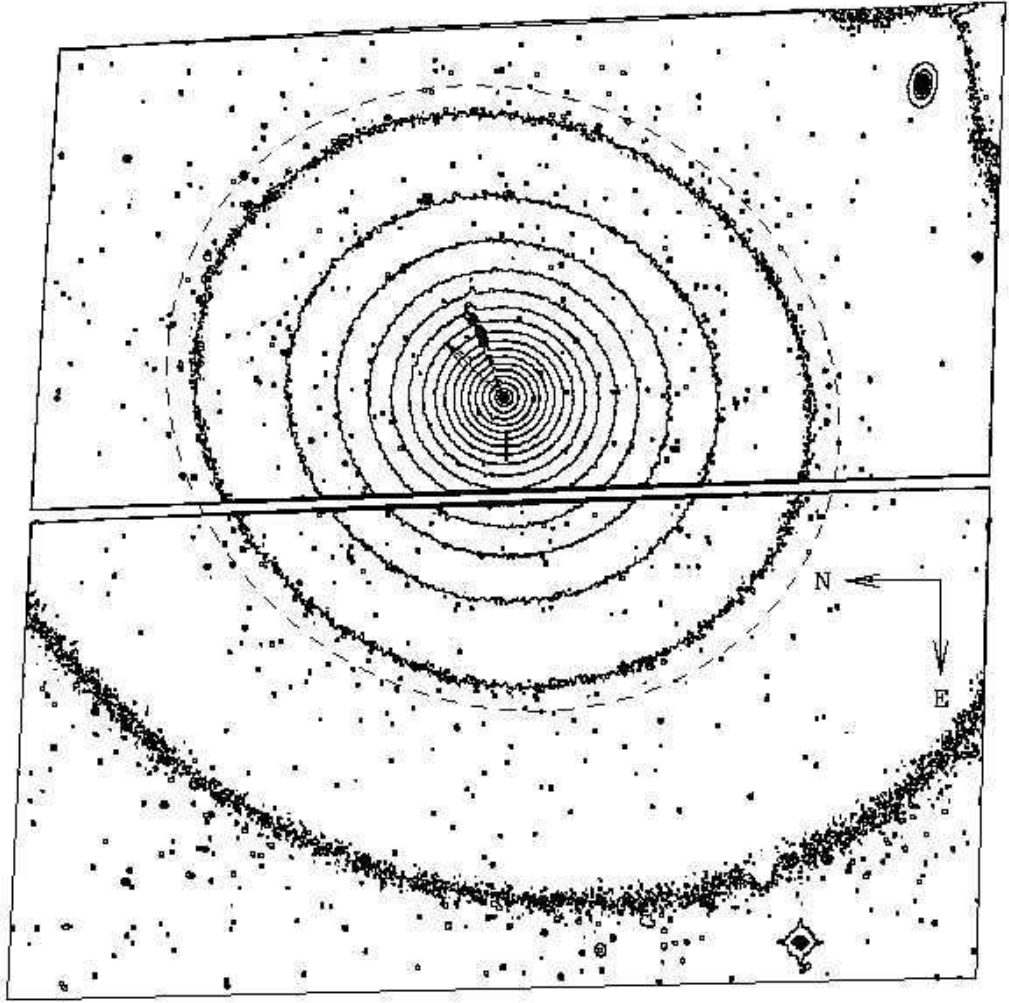}
\caption{Field orientation for VCC1316 (M87). Surface
brightness contours for the co-added, distortion-corrected, background-subtracted
F850LP image are shown by the
thin curves. The cross shows the location of the WFC aperture. The diagonal arrow 
indicates the direction of the WFPC2 parallel field, offset by $\approx 5\farcm8$
in the HST focal plane. The dashed curve shows the ellipse which best-fits the
galaxy isophotes at a surface brightness of $\mu_z = 20$~mag~arcsec$^{-2}$.
\label{fig17}}
\end{figure}

Figures~\ref{fig09}--\ref{fig15} show Digitized Sky Survey
images for each of the 100 program galaxies, ranked in order
of decreasing blue magnitude. In each panel, an overlay shows 
the location and orientation of the ACS/WFC field. Final 
F475W and F850LP images for one program galaxy (VCC1316)
are presented in Figure~\ref{fig16}, while Figure~\ref{fig17}
shows a contour plot for this same galaxy based on the
final co-added, background-subtracted F850LP image.
The heavy cross marks the location of the WFC aperture,
while the arrow shows the direction of the WFPC2 parallel field.
For reference, the dashed curve shows the best-fit ellipse
at a surface brightness of $\mu_z = 20$~mag~arcsec$^{-2}$.

\section{Parallel Observations}
\label{sec:parallel}

The ACS Virgo Cluster Survey includes a significant coordinated
parallel component, using WFPC2 to target 100 ``blank" fields
in Virgo. Since the offset between the
WFPC2 and ACS/WFC cameras in the HST focal plane is $\approx$
5\farcm8 --- corresponding to a projected distance of about 29 kpc
in Virgo --- these fields are effectively ``intergalactic" 
for all but the most luminous galaxies ($i.e.,$ the cluster
potential exceeds that of the adjacent galaxy).

Deep images in these blank fields are useful for two reasons.
First, while the ACS images target the central regions of the 
program galaxies where the
surface density of GCs is highest, accurate 
measurements of the background surface density of GC 
contaminants are required for both
the statistical removal of background sources from the
resulting GC catalogs (a crucial issue for the faintest
galaxies since they contain relatively few GCs)
and the determination of GC surface density 
profiles (see C\^ot\'e et al. 2001 for a demonstration of the 
problems involved).
Second, recent studies of isolated red giant branch stars
and planetary nebulae in Virgo (see Durrell et al. 2002 and
references therein) suggest that 10--20\% of the cluster's
total luminosity may reside in a diffuse, intergalactic
component. Although red giant branch stars and planetary nebulae in Virgo
are much too faint to detect in a single orbit with HST, the
detection of any intergalactic GCs (West et al. 1995;
Jord\'an et al. 2003) associated with this diffuse component
would be straightforward.\footnote{We may estimate the number
of intergalactic GCs as follows.
Durrell et al. (2002) find from their study of intergalactic RGB
stars that $\approx$ 15\% of Virgo's total luminosity resides in
this diffuse component. According to Sandage et al. (19854), the 
luminosity within a region of radius $6^{\circ}$ is
$L_V = 3.3\times10^{12}~L_{V,\odot}$, where we have
scaled their estimate to a distance of 17~Mpc and assumed 
a mean color of $(B-V) = 0.9$ for the intergalactic light.
The luminosity contained within our WFPC2 survey area of 0.16~deg$^2$
is then $7\times10^8~L_{V,\odot}$. Assuming that
we probe the upper 90\% of the GCLF and that the diffuse light
is characterized by a GC specific frequency of $S_N = 4$, we
expect our parallel fields to contain $\sim$ 30 intergalactic GCs.}
Finally, spectroscopic surveys of the Fornax Cluster have
produced a number of candidate ``ultra-compact dwarfs"
($e.g.,$ Phillips et al. 2001; Drinkwater et al. 2003).
The depth and high spatial resolution of the WFPC2 images
will allow the detection of any such ultra-compact dwarfs which
might be present in the Virgo blank fields, and the measurement 
of their properties ($i.e.$, magnitudes, colors and
structural parameters).

While any telescope roll angle would suffice for the ACS/WFC
imaging, the precise value is important for the parallel
observations. For each galaxy, the Visual Target Tuner was
used to align the WFPC2 on regions
of the sky free relatively from bright galaxies and/or stars.
It is especially important to avoid the
former contaminants if one wishes to draw conclusions on the 
properties of putative intergalactic GCs in Virgo.
Images were obtained in two filters, F606W ($\approx V$) 
and F814W ($\approx I$), having respective exposure
times of 700s and 1200s (CR-SPLIT to facilitate the
rejection of cosmic rays). For a point-source signal-to-noise
ratio of S/N = 10, the expected limiting magnitudes for
these images are $V_{\rm lim} \simeq 26.0$ and $I_{\rm lim} \simeq 24.7$.
Since these limiting magnitudes are $\Delta V \simeq 2.1$~mag and
$\Delta I \simeq 1.9$~mag past the $V$- and $I$-band
GCLF turnovers in Virgo (Larsen et al. 2001), the parallel
observations probe the upper 90--95\% of the GCLF in Virgo.
More details on the WFPC2 observations and their analysis
will be presented in a future paper.

\section{Long-slit Spectroscopy}
\label{sec:spectroscopy}

The overarching goal of the ACS Virgo Cluster Survey is an improved
understanding of how early-type galaxies form and evolve. To realize the full
potential of the HST dataset, it is important to utilize constraints
on the dynamics of the galaxies themselves, particularly
the internal distribution of gravitating mass. Since the chemical 
enrichment histories of each galaxy's GC system and underlying
stellar populations likely depend on the depth and shape of its
gravitational potential well, we initiated a ground-based program to obtain
long-slit, integrated-light spectra for each galaxy in the survey.
During the 2003 and 2004 observing seasons, spectra were obtained for 
95 of the 100 program galaxies using three different instrument/telescope
combinations. First, spectra in the range 4000--6000 \AA\ (at an
instrumental resolution of $\sigma_{ins} \approx 60$~km~s$^{-1}$)
were obtained for galaxies with $B_T \le 11.8$ using the GoldCam
spectrometer on the KPNO 2.1m telescope. Second, galaxies in the range 
$11.8 < B_T < 14.6$ were observed with the RC spectrograph
on the KPNO 4m telescope; two grating configurations were
used for these observations, yielding spectra in the range
4200--6200 \AA\ ($\sigma \approx 60$~km~s$^{-1}$) or
4800--5600 \AA\ ($\sigma \approx 35$~km~s$^{-1}$).
Finally, galaxies in the range $14.6 \le B_T \le 16.0$ were
observed using the Echelle Spectrograph and Imager (ESI) on the
Keck II 10m telescope; these spectra cover the wavelength 
range 4000 $\lesssim \lambda \lesssim 10,000$ \AA\ with a 
resolution of $\sigma \approx$ 30~km~s$^{-1}$. A complete description
of these spectroscopic observations, including the determination of
velocity dispersion profiles, rotation curves, mass profiles and
line index measurements, will be presented in future papers.

\section{Summary}
\label{sec:summary}

This paper has provided a brief introduction to the ACS Virgo
Cluster Survey (GO-9401). In terms of depth, spatial
resolution, sample size and homogeneity, this represents the
most comprehensive imaging survey to date of early-type galaxies
in a cluster environment. Scientific results from the survey
will be presented in future papers in this series. More
information on the survey, including object catalogs and
other data products, are available from the program website:
{\sf http://www.physics.rutgers.edu/$\sim$pcote/acs}.

\acknowledgments

Support for program GO-9401 was provided through a grant from the Space
Telescope Science Institute, which is operated by the Association of 
Universities for Research in Astronomy, Inc., under NASA contract 
NAS5-26555. 
P.C. acknowledges additional support provided by NASA LTSA grant NAG5-11714.
A.J. acknowledges additional financial support provided by the National
Science Foundation through a grant from the Association of Universities 
for Research in Astronomy, Inc., under NSF cooperative agreement 
AST-9613615, and by Fundaci\'on Andes under project No.C-13442.
M.M. acknowledges additional financial support provided by the Sherman
M. Fairchild foundation. D.M. is supported by NSF grant AST-020631, 
NASA grant NAG5-9046, and grant HST-AR-09519.01-A from STScI. 
M.J.W. acknowledges support through NSF grant AST-0205960.
This research has made use of the NASA/IPAC Extragalactic Database (NED)
which is operated by the Jet Propulsion Laboratory, California Institute
of Technology, under contract with the National Aeronautics and Space Administration.

\clearpage

\clearpage

\clearpage

\clearpage

\clearpage

\clearpage

\clearpage

\clearpage

\clearpage

\clearpage

\clearpage

\clearpage

\clearpage

\clearpage

\clearpage

\clearpage

\clearpage

\clearpage

\begin{deluxetable}{crccccll}
\tabletypesize{\scriptsize}
\tablecaption{Basic Data for ACS Virgo Cluster Survey Galaxies.\label{tab1}}
\tablewidth{0pt}
\tablehead{
\colhead{ID} & 
\colhead{VCC} & 
\colhead{$\alpha$} & 
\colhead{$\delta$} &
\colhead{$B_T$} &
\colhead{$\langle v_r\rangle$} &
\colhead{Type} &
\colhead{Other} \\
\colhead{} & 
\colhead{} & 
\colhead{(J2000)} & 
\colhead{(J2000)} &
\colhead{(mag)} &
\colhead{(km~s$^{-1}$)} &
\colhead{} &
\colhead{}  
}
\startdata
  1 & 1226 & 12:29:46.79 & +08:00:01.5 &  9.31 &  997$\pm$7 & E2/S0$_1$(2)  & M49,~N4472  \\
  2 & 1316 & 12:30:49.42 & +12:23:28.0 &  9.58 & 1307$\pm$7 & E0            & M87,~N4486  \\
  3 & 1978 & 12:43:39.66 & +11:33:09.4 &  9.81 & 1117$\pm$6 & S0$_1$(2)     & M60,~N4649  \\
  4 &  881 & 12:26:11.74 & +12:56:46.4 & 10.06 & -244$\pm$5 & S0$_1$(3)/E3  & M86,~N4406  \\
  5 &  798 & 12:25:24.04 & +18:11:25.9 & 10.09 &  729$\pm$2 & S0$_1$(3)~pec & M85,~N4382  \\
  6 &  763 & 12:25:03.72 & +12:53:13.2 & 10.26 & 1060$\pm$6 & E1            & M84,~N4374  \\
  7 &  731 & 12:24:28.20 & +07:19:03.0 & 10.51 & 1243$\pm$6 & E3            & N4365  \\
  8 & 1535 & 12:34:03.10 & +07:41:59.0 & 10.61 &  448$\pm$8 & S0$_3$(6)     & N4526  \\
  9 & 1903 & 12:42:02.40 & +11:38:48.0 & 10.76 &  410$\pm$6 & E4            & M59,~N4621  \\
 10 & 1632 & 12:35:39.82 & +12:33:22.6 & 10.78 &  340$\pm$4 & S0$_1$(0)     & M89,~N4552  \\
 11 & 1231 & 12:29:48.87 & +13:25:45.7 & 11.10 & 2244$\pm$2 & E5            & N4473  \\
 12 & 2095 & 12:52:56.00 & +11:13:53.0 & 11.18 &  984$\pm$41& S0$_1$(9)     & N4762  \\
 13 & 1154 & 12:29:00.03 & +13:58:42.9 & 11.37 & 1210$\pm$16& S0$_3$(2)     & N4459  \\
 14 & 1062 & 12:28:03.90 & +09:48:14.0 & 11.40 &  532$\pm$8 & SB0$_1$(6)    & N4442  \\
 15 & 2092 & 12:52:17.50 & +11:18:50.0 & 11.51 & 1377$\pm$15& SB0$_1$(5)    & N4754  \\
 16 &  369 & 12:19:45.42 & +12:47:54.3 & 11.80 & 1009$\pm$13& SB0$_1$       & N4267  \\
 17 &  759 & 12:24:55.50 & +11:42:15.0 & 11.80 &  943$\pm$19& SB0$_2$(r)(3) & N4371  \\
 18 & 1692 & 12:36:53.40 & +07:14:47.0 & 11.82 & 1730$\pm$13& S0$_1$(7)/E7  & N4570  \\
 19 & 1030 & 12:27:40.49 & +13:04:44.2 & 11.84 &  801$\pm$10& SB0$_1$(6)    & N4435  \\
 20 & 2000 & 12:44:31.95 & +11:11:25.1 & 11.94 & 1083$\pm$4 & E3/S0$_1$(3)  & N4660  \\
 21 &  685 & 12:23:57.90 & +16:41:37.0 & 11.99 & 1241$\pm$19& S0$_1$(8)     & N4350  \\
 22 & 1664 & 12:36:26.86 & +11:26:20.6 & 12.02 & 1142$\pm$2 & E6            & N4564  \\
 23 &  654 & 12:23:35.28 & +16:43:22.3 & 12.03 &  950$\pm$9 & RSB0$_2$(5)   & N4340  \\
 24 &  944 & 12:26:50.53 & +09:35:02.0 & 12.08 &  843$\pm$15& S0$_1$(7)     & N4417  \\
 25 & 1938 & 12:42:47.40 & +11:26:33.0 & 12.11 & 1164$\pm$10& S0$_1$(7)     & N4638  \\
 26 & 1279 & 12:30:17.39 & +12:19:43.9 & 12.15 & 1349$\pm$3 & E2            & N4478  \\
 27 & 1720 & 12:37:30.61 & +09:33:18.8 & 12.29 & 2273$\pm$12& S0$_{1/2}$(4) & N4578  \\
 28 &  355 & 12:19:30.61 & +14:52:41.4 & 12.41 & 1359$\pm$4 & SB0$_{2/3}$   & N4262  \\
 29 & 1619 & 12:35:30.61 & +12:13:15.4 & 12.50 &  381$\pm$9 & E7/S0$_1$(7)  & N4550  \\
 30 & 1883 & 12:41:32.70 & +07:18:53.0 & 12.57 & 1875$\pm$22& RSB0$_{1/2}$  & N4612  \\
 31 & 1242 & 12:29:53.49 & +14:04:07.0 & 12.60 & 1609$\pm$11& S0$_1$(8)     & N4474  \\
 32 &  784 & 12:25:14.75 & +15:36:27.2 & 12.67 & 1069$\pm$10& S0$_1$(2)     & N4379  \\
 33 & 1537 & 12:34:06.10 & +11:19:17.0 & 12.70 & 1374$\pm$10& SB0$_2$(5)    & N4528  \\
 34 &  778 & 12:25:12.27 & +14:45:43.8 & 12.72 & 1375$\pm$11& S0$_1$(3)     & N4377  \\
 35 & 1321 & 12:30:52.21 & +16:45:32.6 & 12.84 &  967$\pm$6 & S0$_1$(1)     & N4489  \\
 36 &  828 & 12:25:41.70 & +12:48:38.0 & 12.84 &  561$\pm$15& E5            & N4387  \\
 37 & 1250 & 12:29:59.10 & +12:20:55.0 & 12.91 & 1970$\pm$11& S0$_3$(5)     & N4476  \\
 38 & 1630 & 12:35:37.97 & +12:15:50.5 & 12.91 & 1172$\pm$6 & E2            & N4551  \\
 39 & 1146 & 12:28:57.56 & +13:14:30.8 & 12.93 &  635$\pm$6 & E1            & N4458  \\
 40 & 1025 & 12:27:36.71 & +08:09:14.8 & 13.06 & 1071$\pm$6 & E0/S0$_1$(0)  & N4434  \\
 41 & 1303 & 12:30:40.64 & +09:00:55.9 & 13.10 &  875$\pm$10& SB0$_1$(5)    & N4483  \\
 42 & 1913 & 12:42:10.70 & +07:40:37.0 & 13.22 & 1892$\pm$37& E7            & N4623  \\
 43 & 1327 & 12:30:57.56 & +12:16:17.2 & 13.26 &  150$\pm$45& E2            & N4486A \\
 44 & 1125 & 12:28:43.37 & +11:45:21.0 & 13.30 &  195$\pm$47& S0$_1$(9)     & N4452  \\
 45 & 1475 & 12:33:04.95 & +16:15:55.9 & 13.36 &  951$\pm$11& E2            & N4515  \\
 46 & 1178 & 12:29:21.30 & +08:09:23.0 & 13.37 & 1243$\pm$2 & E3            & N4464  \\
 47 & 1283 & 12:30:18.40 & +13:34:40.9 & 13.45 &  876$\pm$10& SB0$_2$(2)    & N4479  \\
 48 & 1261 & 12:30:10.39 & +10:46:46.1 & 13.56 & 1871$\pm$16& d:E5,N        & N4482  \\
 49 &  698 & 12:24:05.00 & +11:13:06.0 & 13.60 & 2080$\pm$10& S0$_1$(8)     & N4352  \\
 50 & 1422 & 12:32:14.21 & +10:15:05.0 & 13.64 & 1288$\pm$10& E1,N:         & I3468  \\
 51 & 2048 & 12:47:15.32 & +10:12:13.0 & 13.81 & 1084$\pm$12& d:S0(9)       & I3773  \\
 52 & 1871 & 12:41:15.72 & +11:23:13.5 & 13.86 &  567$\pm$10& E3            & I3653  \\
 53 &    9 & 12:09:22.34 & +13:59:33.1 & 13.93 & 1804$\pm$49& dE1,N         & I3019  \\
 54 &  575 & 12:22:43.31 & +08:11:53.7 & 14.14 & 1231$\pm$9 & E4            & N4318  \\
 55 & 1910 & 12:42:08.69 & +11:45:14.9 & 14.17 &  206$\pm$26& dE1,N         & I809   \\
 56 & 1049 & 12:27:54.86 & +08:05:25.2 & 14.20 &  716$\pm$36& S0(4)         & U7580  \\
 57 &  856 & 12:25:57.81 & +10:03:12.8 & 14.25 & 1025$\pm$10& dE1,N         & I3328  \\
 58 &  140 & 12:15:12.58 & +14:25:59.1 & 14.30 & 1072$\pm$46& S0$_{1/2}$(4) & I3065  \\
 59 & 1355 & 12:31:20.04 & +14:06:53.5 & 14.31 & 1332$\pm$63\tablenotemark{a} & dE2,N         & I3442  \\
 60 & 1087 & 12:28:14.90 & +11:47:24.0 & 14.31 &  675$\pm$12& dE3,N         & I3381  \\
 61 & 1297 & 12:30:31.85 & +12:29:26.0 & 14.33 & 1555$\pm$4 & E1            & N4486B \\
 62 & 1861 & 12:40:58.54 & +11:11:04.4 & 14.37 &  470$\pm$39& dE0,N         & I3652  \\
 63 &  543 & 12:22:19.55 & +14:45:38.6 & 14.39 &  985$\pm$12& dE5           & U7436  \\
 64 & 1431 & 12:32:23.37 & +11:15:46.2 & 14.51 & 1500$\pm$38& dE0,N         & I3470  \\
 65 & 1528 & 12:33:51.62 & +13:19:21.3 & 14.51 & 1608$\pm$35& d:E1          & I3501  \\
 66 & 1695 & 12:36:54.87 & +12:31:12.5 & 14.53 & 1547$\pm$29& dS0:          & I3586  \\
 67 & 1833 & 12:40:19.65 & +15:56:07.2 & 14.54 & 1679$\pm$34& S0$_1$(6)     &        \\
 68 &  437 & 12:20:48.82 & +17:29:13.4 & 14.54 & 1474$\pm$46& dE5,N         & U7399A \\
 69 & 2019 & 12:45:20.42 & +13:41:33.0 & 14.55 & 1895$\pm$44& dE4,N         & I3735  \\
 70 &   33 & 12:11:07.76 & +14:16:29.8 & 14.67 & 1093$\pm$52& d:E2,N:       & I3032  \\
 71 &  200 & 12:16:33.68 & +13:01:53.1 & 14.69 &   65$\pm$43& dE2,N         &        \\
 72 &  571 & 12:22:41.14 & +07:57:01.1 & 14.74 & 1047$\pm$37& SB0$_1$(6)    &        \\
 73 &   21 & 12:10:23.19 & +10:11:17.6 & 14.75 &  506$\pm$35& dS0(4)        & I3025  \\
 74 & 1488 & 12:33:13.44 & +09:23:49.8 & 14.76 & 1157$\pm$48& E6:           & I3487  \\
 75 & 1779 & 12:39:04.67 & +14:43:51.5 & 14.83 & 1313$\pm$45& dS0(6):       & I3612  \\
 76 & 1895 & 12:41:52.00 & +09:24:10.3 & 14.91 & 1032$\pm$51& d:E6          & U7854  \\
 77 & 1499 & 12:33:19.79 & +12:51:12.8 & 14.94 & -575$\pm$35& E3~pec~or~S0  & I3492  \\
 78 & 1545 & 12:34:11.54 & +12:02:55.9 & 14.96 & 2050$\pm$32& E4            & I3509  \\
 79 & 1192 & 12:29:30.20 & +07:59:34.0 & 15.04 & 1426$\pm$22& E3            & N4467  \\
 80 & 1857 & 12:40:53.10 & +10:28:34.0 & 15.07 &  634$\pm$69& dE4:,N?       & I3647  \\
 81 & 1075 & 12:28:12.29 & +10:17:51.0 & 15.08 & 1844$\pm$40& dE4,N         & I3383  \\
 82 & 1948 & 12:42:58.02 & +10:40:54.5 & 15.10 & 1672$\pm$98& dE3           &        \\
 83 & 1627 & 12:35:37.25 & +12:22:54.9 & 15.16 &  236$\pm$41& E0            &        \\
 84 & 1440 & 12:32:33.39 & +15:24:55.2 & 15.20 &  414$\pm$44& E0            & I798   \\
 85 &  230 & 12:17:19.64 & +11:56:36.2 & 15.20 & 1490$\pm$65& dE4:,N:       & I3101  \\
 86 & 2050 & 12:47:20.69 & +12:09:58.7 & 15.20 & 1193$\pm$48& dE5:,N        & I3779  \\
 87 & 1993 & 12:44:12.02 & +12:56:30.1 & 15.30 &  875$\pm$50& E0            &        \\
 88 &  751 & 12:24:48.34 & +18:11:42.0 & 15.30 &  710$\pm$39& dS0           & I3292  \\
 89 & 1828 & 12:40:13.38 & +12:52:29.0 & 15.33 & 1517$\pm$57& dE2,N         & I3635  \\
 90 &  538 & 12:22:14.83 & +07:10:00.8 & 15.40 &  500$\pm$50& E0            & N4309A \\
 91 & 1407 & 12:32:02.69 & +11:53:24.8 & 15.49 & 1001$\pm$11& dE2,N         & I3461  \\
 92 & 1886 & 12:41:39.41 & +12:14:52.4 & 15.49 & 1159$\pm$65& dE5,N         &        \\
 93 & 1199 & 12:29:34.97 & +08:03:31.4 & 15.50 &  900$\pm$50& E2            &        \\
 94 & 1743 & 12:38:06.77 & +10:04:56.6 & 15.50 & 1279$\pm$10& dE6           & I3602  \\
 95 & 1539 & 12:34:06.77 & +12:44:30.1 & 15.68 & 1390$\pm$50& dE0,N         &        \\
 96 & 1185 & 12:29:23.43 & +12:27:02.4 & 15.68 &  500$\pm$50& dE1,N         &        \\
 97 & 1826 & 12:40:11.24 & +09:53:45.9 & 15.70 & 2033$\pm$38& dE2,N         & I3633  \\
 98 & 1512 & 12:33:34.56 & +11:15:42.8 & 15.73 &  762$\pm$35& dS0~pec       &        \\
 99 & 1489 & 12:33:13.84 & +10:55:43.6 & 15.89 &   80$\pm$50& dE5,N?        & I3490  \\
100 & 1661 & 12:36:24.81 & +10:23:04.6 & 15.97 & 1400$\pm$50& dE0,N         &        \\
\enddata

\tablenotetext{a}{ Velocity from Binggeli, Popescu \& Tammann (1993). Note that
Karachentsev \& Karachentsev (1982) claim a (heliocentric) velocity of
$\langle v_r \rangle = 6210\pm60$ km~s$^{-1}$.}

\tablecomments{Units of right ascension are hours, minutes, and seconds, 
and units of declination are degrees, arcminutes, and arcseconds.}

\end{deluxetable}

\clearpage

\begin{deluxetable}{lc}
\tablecaption{Morphologies of ACS Virgo Cluster Survey Galaxies.\label{tab2}}
\tablewidth{0pt}
\tablehead{
\colhead{Type} &
\colhead{Number} 
}
\startdata
E                     &  26 \\
E/S0                  &   7 \\
S0                    &  32 \\
dE                    &   5 \\
dE,N                  &  24 \\
dS0                   &   6 \\
 \cline{1-2} 
All                   & 100 \\
\enddata
\end{deluxetable}

\clearpage

\end{document}